\documentclass[aps,pre,preprint]{revtex4}
\usepackage{amsmath}
\usepackage{graphics}
\newtheorem{lem}{Lemma}
\newtheorem{defi}{Definition}

\begin{document}
\baselineskip = 0.54cm
\parbox{0.85\textwidth}{\Large\bf General soliton matrices in the Riemann-Hilbert problem
for integrable nonlinear equations }\\
\begin{center}
\parbox{0.8\textwidth}{{\sf  Valery S. Shchesnovich}${}^{a)}$
{\sf and Jianke Yang}${}^{b)}$\\
{\it Department of Mathematics and Statistics,
University of Vermont, Burlington VT 05401, USA}\\[0.5cm]

\baselineskip=0.54cm

We derive the soliton matrices corresponding to an arbitrary number of higher-order
normal zeros for the matrix Riemann-Hilbert problem of arbitrary matrix dimension,
thus giving the complete solution to the problem of higher-order solitons. Our
soliton matrices explicitly give all higher-order multi-soliton solutions to the
nonlinear partial differential equations integrable through the matrix
Riemann-Hilbert problem. We have applied these general results to the three-wave
interaction system, and derived new classes of higher-order soliton and two-soliton
solutions, in complement to those from our previous publication [Stud. Appl. Math.
\textbf{110}, 297 (2003)], where only the elementary higher-order zeros were
considered. The higher-order solitons corresponding to non-elementary zeros
generically describe the simultaneous breakup of a pumping wave $(u_3)$ into the
other two components ($u_1$ and $u_2$) and merger of $u_1$ and $u_2$ waves into the
pumping $u_3$ wave. The two-soliton solutions corresponding to two simple zeros
generically describe the breakup of the pumping $u_3$ wave into the $u_1$ and $u_2$
components, and the reverse process. In the non-generic cases, these two-soliton
solutions could describe the elastic interaction of the $u_1$ and $u_2$ waves, thus
reproducing previous results obtained by Zakharov and Manakov [Zh. Eksp. Teor. Fiz.
\textbf{69}, 1654 (1975)] and Kaup [Stud. Appl. Math. \textbf{55}, 9 (1976)]. }
\end{center}

\vspace{1cm} Keywords: matrix Riemann-Hilbert problem;  soliton
solutions to integrable nonlinear PDEs.

\vfill \hrule
\medskip

\noindent
{{\small\rm  ${}^{a)}$ Instituto de F{i}sica Te\'{o}rica, Universidade Estadual Paulista,
 Rua Pamplona 145, 01405-900 S\~{a}o Paulo, Brazil\\
 Email: valery@ift.unesp.br}
\medskip

\noindent
{\small\rm ${}^{b)}$ Email: jyang@emba.uvm.edu}
\newpage

\section{Introduction}

The importance of integrable nonlinear partial differential equations (PDEs) in 1+1
dimensions in applications to nonlinear physics can hardly be overestimated. Their
importance partially stems from the fact that it is always possible to obtain
certain explicit solutions, called solitons, by some algebraic procedure. At
present, there is a wide range of literature concerning integrable nonlinear PDEs
and their soliton solutions (see, for instance, Refs.~\cite{AS81,NMPZ84,FT87,AC91}
and the references therein). The reader familiar with the inverse scattering
transform method knows that it is zeros of the Riemann-Hilbert problem (or poles of
the reflection coefficients in the previous nomenclature) that give rise to the soliton
solutions. These solutions are usually derived by using one of the several well-known
techniques, such as the dressing method \cite{AS81,ZS79a,ZS79b}, the
Riemann-Hilbert problem approach \cite{NMPZ84,FT87}, and the Hirota method (see \cite{AS81}).
In the first two methods,
the pure soliton solution is obtained by considering the asymptotic form of
a rational matrix function of the spectral parameter, called the soliton matrix in
the following. It is known, that the generic case of zeros of the matrix
Riemann-Hilbert problem is the case of simple zeros
\cite{BC1,BC2,BC3,BDT,Zhou1,Zhou2} (see also Ref.~\cite{Kawata}). A single simple
zero produces a one-soliton solution. Several distinct zeros will produce
multi-soliton solutions, which describe the interaction (scattering) of individual
solitons. As far as the generic case is concerned, there is no problem in
the derivation of the corresponding soliton solutions.

However, in the non-generic cases, when at least one higher-order (i.e. multiple)
zero is present in the Riemann-Hilbert problem, the situation is not so definite.
Higher-order zeros must be considered separately, as, in general, the soliton
solutions which correspond to such zeros cannot be derived from the known generic
multi-soliton solutions by coalescing some of the distinct simple zeros. This is
clear from the fact that a higher-order zero generally corresponds to a
higher-order pole in the soliton matrix (or its inverse), which cannot be obtained
in a regular way by coalescing simple poles in the generic multi-soliton matrix.
The procedure of coalescing several distinct simple zeros produces only
higher-order zeros with equal algebraic and geometric multiplicities (the geometric
multiplicity is defined as the dimension of the kernel of the soliton matrix
evaluated at the zero), which is just the trivial case of higher-order zeros. For
instance, if the algebraic multiplicity is equal or greater than the matrix
dimension, then such coalescing will produce a higher-order zero with the geometric
multiplicity no less than the matrix dimension, which could only correspond to the
zero solution instead of solitons. Thus the soliton matrices corresponding to the
higher-order zeros of the Riemann-Hilbert problem require a separate consideration.

Soliton solutions corresponding to higher-order zeros have been investigated in the
literature before, mainly for the $2\times2$-dimensional spectral problem. A
soliton solution to the nonlinear Schr\"odinger (NLS) equation corresponding to a
double zero was first given in Ref.~\cite{ZS72} but without much analysis. The
double- and triple-zero soliton solutions to the KdV equation were examined in
Ref.~\cite{Wadati} and the general multiple-zero soliton solution to the sine-Gordon
equation was extensively studied in Ref.~\cite{Tsuru} using the associated
Gelfand-Levitan-Marchenko equation. In Refs.~\cite{OptLett,Nathalie}, higher-order
soliton solutions to the NLS equation were studied by employing the dressing
method. In \cite{pelinovsky,ablowitz,ablowitz2}, higher order solitons in the
Kadomtsev-Petviashvili~I equation were derived by the direct method and
the inverse scattering method.
Finally, in our previous publication \cite{SIAM} we have derived soliton
matrices corresponding to a single {\em elementary} higher-order zero --- a zero
which has the geometric multiplicity equal to 1. Our studies give the general
higher-order soliton solutions for the integrable
PDEs associated with the $2\times2$ matrix Riemann-Hilbert problem with
a single higher-order zero. Indeed, any
zero of the $2\times2$-dimensional Riemann-Hilbert problem is elementary since a
nonzero $2\times2$ matrix can have only one vector in its kernel.

However, the previous investigations left some of the key questions unanswered. For
instance, the general soliton matrix corresponding to a single non-elementary zero
remained unknown. Such zeros arise when the matrix dimension of the Riemann-Hilbert
problem is greater than 2. Naturally then, the ultimate question --- the most
general soliton matrices corresponding to an arbitrary number of higher-order zeros
in the general $N\times N$ Riemann-Hilbert problem, was not addressed. Because of
these unresolved issues, the most general soliton and multi-soliton solutions to
PDEs integrable through the $N\times N$ Riemann-Hilbert problem (such as the NLS
equation \cite{zakharov}, the three-wave interaction system
\cite{NMPZ84,3wave1,3wave2,3wavezakharov,3wave3}, and the Manakov equations
\cite{manakov}) have not been derived yet.

In this paper we derive the complete solution to the problem of soliton matrices
corresponding to an arbitrary number of higher-order normal zeros for the general
$N\times N$ matrix Riemann-Hilbert problem. These normal zeros are defined in
Definition \ref{def1}, and are non-elementary in general. They include almost all
physically important integrable PDEs where the involution property [see Eq.
(\ref{invol})] holds. The corresponding soliton solutions can be termed as the
higher-order multi-solitons, to reflect the fact that these solutions do not belong
to the class of the previous generic multi-soliton solutions. Our results give a
complete classification of all possible soliton solutions in the integrable PDEs
associated with the $N\times N$ Riemann-Hilbert problem. In other words, our
soliton matrices contain the most general forms of reflection-less (soliton)
potentials in the $N$-dimensional Zakharov-Shabat spectral operator. For these
general soliton potentials, the corresponding discrete and continuous
eigenfunctions of the $N$-dimensional Zakharov-Shabat operator naturally follow
from our soliton matrices. As an example, we consider  the three-wave interaction
system, and derive single-soliton solutions corresponding to a non-elementary zero,
and higher-order two-soliton solutions. These solutions generate many new processes
such as the simultaneous breakup of a pumping wave $(u_3)$ into the other two
components ($u_1$ and $u_2$) and merger of $u_1$ and $u_2$ waves into the pumping
$u_3$ wave, i.e., $u_1+u_2+u_3 \leftrightarrow u_1+u_2+u_3$. They also reproduce
previous solitons in \cite{NMPZ84,3wavezakharov,3wave3,SIAM} as special cases.

The paper is organized as follows. A summary on the Riemann-Hilbert problem is
placed in section \ref{secRH}. Section \ref{multpl} is the central section of the
paper. There we present the theory of soliton matrices corresponding to several
higher-order zeros under the assumption that these zeros are {\em normal} (see
Definition \ref{def1}), which include the physically important cases
with the involution property [see Eq. (\ref{invol})].
Applications of these general results to the three-wave interaction
system are contained in Section \ref{sec3wave}. Finally, in the appendix we briefly treat
the more general case where the zeros are abnormal.

\section{The Riemann-Hilbert problem approach}
\label{secRH}

The integrable nonlinear PDEs in 1+1 dimensions are associated
with the matrix Riemann-Hilbert problem (consult, for instance,
Refs.
\cite{AS81,NMPZ84,FT87,AC91,ZS79a,ZS79b,Fokas1,Fokas2,Fokas3,Leon,BC1,BC2,BC3,BDT,Zhou1,Zhou2}).
The matrix Riemann-Hilbert problem (below we work in the space of
$N\times N$ matrices) is the problem of finding the holomorphic
factorization, denoted below by  $\Phi_+(k)$ and $\Phi^{-1}_-(k)$,
in the complex plane of  a nondegenerate matrix function $G(k)$
given on an oriented curve~$\gamma$:
\begin{equation}
\Phi_-^{-1}(k,x,t)\Phi_+(k,x,t) = G(k,x,t)\equiv
E(k,x,t)G(k,0,0)E^{-1}(k,x,t),\quad k\in\gamma,
\label{RH1}\end{equation}
where
\[
E(k,x,t)\equiv\exp{[-\Lambda(k)x-\Omega(k)t]}.
\]
Here the matrix functions $\Phi_+(k)$ and $\Phi_-^{-1}(k)$ are holomorphic in the
two complementary domains of the complex $k$-plane: $C_+$ to the left and $C_-$ to
the right from the curve $\gamma$, respectively. The matrices $\Lambda(k)$ and
$\Omega(k)$ are called the dispersion laws. Usually the dispersion laws commute
with each other, e.g., given by diagonal matrices. We will consider this case
(precisely in this case $E(k,x,t)$ is given by the above formula). The
Riemann-Hilbert problem requires an appropriate normalization condition. Usually
the curve $\gamma$ contains the infinite point $k=\infty$ of the complex plane and
the normalization condition is formulated as
\begin{equation}
\Phi_\pm(k,x,t) \to I,\quad \text{as}\quad k\to \infty.
\label{RH2}\end{equation}
This normalization condition is called the canonical normalization.
Setting the normalization condition to an arbitrary nondegenerate
matrix function $S(x,t)$ leads to the gauge equivalent integrable
nonlinear PDE, e.g., the Landau-Lifshitz equation in the case of the
NLS equation \cite{FT87}. Obviously, the new solution
$\hat{\Phi}_\pm(k,x,t)$  to the Riemann-Hilbert problem, normalized to
$S(x,t)$, is related to the canonical solution by the following
transformation
\begin{equation}
\hat{\Phi}_\pm(k,x,t) = S(x,t)\Phi(k,x,t).
\label{gaugePhi}\end{equation}
Thus, without any loss of generality, we confine ourselves to the
Riemann-Hilbert problem under the canonical normalization.

For physically applicable nonlinear PDEs the Riemann-Hilbert problem possesses the
involution properties, which reduce the number of the dependent variables (complex
fields). The following involution property of the Riemann-Hilbert problem is the
most common in applications
\begin{equation}
\Phi_+^\dag(k) = \Phi_-^{-1}(\overline{k}),\quad \overline{k} = k^*.
\label{invol}\end{equation}
Here the superscript ``$\dag$'' represents the Hermitian conjugate, and ``*'' the
complex conjugate. Examples include the NLS equation, the Manakov equations,
and the N-wave system.
The analysis in this article includes this involution (\ref{invol}) as a special case.
In this case, the overline of a quantity represents
its Hermitian conjugation in the case of vectors and
matrices and the complex conjugation in the case of scalar quantities.
In other cases, the original and overlined quantities may not be related.

To solve the Cauchy problem for the integrable nonlinear PDE posed on
the whole axis $x$, one usually constructs the associated
Riemann-Hilbert problem  starting with the linear spectral equation
\begin{equation}
\partial_x \Phi(k,x,t) = \Phi(k,x,t)\Lambda(k) + U(k,x,t)\Phi(k,x,t),
\label{RH3}\end{equation}
whereas the $t$-dependence is given by a similar equation
\begin{equation}
\partial_t \Phi(k,x,t) = \Phi(k,x,t)\Omega(k) + V(k,x,t)\Phi(k,x,t).
\label{RH4}\end{equation}
The nonlinear integrable PDE corresponds to the compatibility condition
of the system (\ref{RH3}) and (\ref{RH4}):
\begin{equation}
\partial_tU-\partial_xV+[U,V]=0.
\label{RH5}\end{equation}

The essence of the approach based on the Riemann-Hilbert problem
lies in the fact that the evolution governed by the complicated
nonlinear PDE (\ref{RH5}) is mapped to the evolution of the
spectral data given by simpler equations such as (\ref{RH1}) and
(\ref{RH8a})-(\ref{RH8b}).
When the spectral data is known,
the matrices $U(k,x,t)$ and $V(k,x,t)$ describing the evolution of
$\Phi_\pm$ can then be retrieved from the Riemann-Hilbert problem. In
our case, the potentials $U(k,x,t)$ and $V(k,x,t)$ are completely
determined by the (diagonal) dispersion laws $\Lambda(k)$ and
$\Omega(k)$ and the Riemann-Hilbert solution
$\Phi\equiv\Phi_\pm(k,x,t)$. Indeed, let us assume that the
dispersion laws are polynomial functions, i.e.,
\begin{equation}
\Lambda(k) = \sum_{j=0}^{J_1}A_jk^j,\quad \Omega(k)=\sum_{j=0}^{J_2}
B_j k^j.
\label{polynom}\end{equation}
Then using similar arguments as in Ref.~\cite{Leon} we get:
\begin{equation}
U = -{\cal P}\{\Phi\Lambda\Phi^{-1}\}, \quad V = -{\cal
P}\{\Phi\Omega\Phi^{-1}\}.
\label{UV}\end{equation}
Here the matrix function $\Phi(k)$ is expanded into the asymptotic
series,
\[
\Phi(k)=I+k^{-1}\Phi^{(1)}+k^{-2}\Phi^{(2)}+..., \quad k\to \infty,
\]
and the operator ${\cal P}$ cuts out the polynomial asymptotics of its
argument as $k\to\infty$. An important property of matrices $U$ and $V$
is that
\begin{equation}
\text{Tr}U(k,x,t) = - \text{Tr}\Lambda(k),\quad \text{Tr}V(k,x,t) = -
\text{Tr}\Omega(k),
\label{Traces}\end{equation}
which evidently follows from equation (\ref{UV}). This property
guarantees that the Riemann-Hilbert zeros are $(x,t)$-independent.

Let us consider as an example the physically relevant three-wave
interaction system \cite{NMPZ84,3wave1,3wave2,3wave3}. Set $N=3$,
\begin{equation}
\Lambda(k) = ikA,\quad A =  \left(\begin{array}{ccc}a_1 & 0 & 0\\
0 & a_2 & 0\\ 0 & 0 & a_3  \end{array}\right),\qquad
\Omega(k) = ikB,\quad B =  \left(\begin{array}{ccc}b_1 & 0 & 0\\
0 & b_2 & 0\\ 0 & 0 & b_3  \end{array}\right),
\label{disper}\end{equation}
where $a_j$ and $b_j$ are real with the elements of $A$  being ordered:
$a_1>a_2>a_3$. From equation (\ref{UV}) we get
\begin{equation}
U = -\Lambda(k) + i[A,\Phi^{(1)}],\quad V = -\Omega(k) +
i[B,\Phi^{(1)}].
\label{UV3wave}\end{equation}
Setting
\begin{equation}
u_1 = \sqrt{a_1-a_2}\Phi^{(1)}_{12},\quad u_2 =
\sqrt{a_2-a_3}\Phi^{(1)}_{23},\quad u_3 =
\sqrt{a_1-a_3}\Phi^{(1)}_{13},
\label{u1u2u3}\end{equation}
assuming the involution (\ref{invol}),  and using equation (\ref{UV3wave}) in
(\ref{RH5}) we get the three-wave system:
\begin{subequations}
\label{3wave}
\begin{eqnarray}
\partial_t u_1 + v_1 \partial_x u_1 +i\varepsilon\overline{u}_2u_3 &=&
0,\\
\partial_t u_2 + v_2 \partial_x u_2 +i\varepsilon\overline{u}_1u_3 &=&
0,\\
\partial_t u_3 + v_3 \partial_x u_3 +i\varepsilon {u}_1u_2 &=& 0.
\end{eqnarray}\end{subequations}
Here
\begin{equation}
v_1 = \frac{b_2-b_1}{a_1-a_2},\quad v_2 = \frac{b_3-b_2}{a_2-a_3},\quad v_3 =
\frac{b_3-b_1}{a_1-a_3},
\label{v123}\end{equation}
\begin{equation}
\varepsilon = \frac{a_1b_2-a_2b_1+a_2b_3-a_3b_2+a_3b_1-a_1b_3}
{[(a_1-a_2)(a_2-a_3)(a_1-a_3)]^{1/2}}.
\label{varepsilon}
\end{equation}
The group velocities satisfy the following condition
\begin{equation}
\frac{v_2-v_3}{v_1-v_3} = -\frac{a_1-a_2}{a_2-a_3}<0.
\label{ineqv}\end{equation}
The three-wave system (\ref{3wave}) can be interpreted physically. It
describes the interaction of three wave packets with complex envelopes
$u_1$, $u_2$ and $u_3$ in a medium with quadratic nonlinearity.

In general, the Riemann-Hilbert problem (\ref{RH1})-(\ref{RH2})
has multiple solutions. Different solutions  are related to each
other by the rational matrix functions $\Gamma(k)$  (which also
depend on the variables $x$ and $t$)
\cite{NMPZ84,FT87,ZS79a,ZS79b,Kawata}:
\begin{equation}
\widetilde{\Phi}_\pm(k,x,t) = \Phi_\pm(k,x,t)\Gamma(k,x,t).
\label{RH6}\end{equation}
The rational matrix $\Gamma(k)$ must satisfy the canonical
normalization condition: $\Gamma(k)\to I$ for $k\to \infty$ and must
have poles only in $C_-$ (the inverse function  $\Gamma^{-1}(k)$ then
has poles in $C_+$ only). Such a rational matrix $\Gamma(k)$ will be
called the soliton matrix below, since it gives the soliton part of the
solution to the integrable nonlinear PDE.

To specify a unique solution to the Riemann-Hilbert problem the set of the
Riemann-Hilbert data must be given. These data are also called the spectral data.
The full set of the spectral data comprises the matrix $G(k,x,t)$ on the right-hand
side of equation (\ref{RH1}) and the appropriate discrete data related to the zeros
of $\det\Phi_+(k)$ and $\det\Phi_-^{-1}(k)$.
In the case of involution (\ref{invol}), the zeros of
$\det\Phi_+(k)$ and $\det\Phi_-^{-1}(k)$ appear in complex conjugate pairs,
$\overline{k}_j = k^*_j$.
It is known
\cite{BC1,BC2,BC3,BDT,Zhou1,Zhou2} (see also Ref.~\cite{Kawata}) that in the
generic case the spectral data include simple (distinct) zeros $k_1,\ldots,k_n$ of
$\det\Phi_+(k)$ and $\overline{k}_1,\ldots,\overline{k}_n$ of $\det\Phi_-^{-1}(k)$,
in their holomorphicity domains, and the null vectors
$|v_1\rangle,\ldots,|v_n\rangle$ and $\langle \overline{v}_1|,\ldots,\langle
\overline{v}_n|$ from the respective kernels:
\begin{equation}
\Phi_+(k_j)|v_j\rangle = 0,\quad \langle
\overline{v}_j|\Phi_-^{-1}(\overline{k}_j)=0.
\label{vects}\end{equation}

Using the property (\ref{Traces}) one can verify that the zeros do not depend
on the variables $x$ and $t$. The  $(x,t)$-dependence of the null vectors can
be easily derived by differentiation of (\ref{vects}) and use of the linear
spectral equations (\ref{RH3})-(\ref{RH4}). This dependence reads:
\begin{subequations}
\label{RH8}
\begin{eqnarray}
|v_j\rangle &=&
\exp{\{-\Lambda(k_j)x-\Omega(k_j)t\}}|v_{0j}\rangle,
\label{RH8a}\\ \langle \overline{v}_j| &=& \langle
\overline{v}_{0j}|\exp{\{\Lambda(\overline{k}_j)x+\Omega(\overline{k}_j)t\}},
\label{RH8b}\end{eqnarray}
\end{subequations}
where $|v_{0j}\rangle$ and $\langle \overline{v}_{0j}|$ are
constant vectors.

The vectors in equations (\ref{RH8a})-(\ref{RH8b}) together with
the zeros constitute the full set of the generic discrete data necessary
to specify the soliton matrix $\Gamma(k,x,t)$ and,  hence, unique
solution to the Riemann-Hilbert problem (\ref{RH1})-(\ref{RH2}).
Indeed, by constructing the soliton matrix $\Gamma(k)$ such that
the following matrix functions
\begin{equation}
\phi_+(k) = \Phi_+(k)\Gamma^{-1}(k),\quad \phi_-^{-1}(k) =
\Gamma(k)\Phi_-^{-1}(k)
\label{regularphi}\end{equation}
are nondegenerate and holomorphic in the domains $C_+$ and $C_-$,
respectively, we reduce the Riemann-Hilbert problem with zeros to
another one without zeros and hence uniquely solvable (for details see,
for instance, Refs.~\cite{NMPZ84,FT87,AC91,Kawata}). Below by matrix
$\Gamma(k)$ we will imply the matrix from equation (\ref{regularphi})
which reduces the Riemann-Hilbert problem (\ref{RH1})-(\ref{RH2}) to
the one without zeros. The corresponding  solution to the integrable
PDE (\ref{RH5}) is obtained by using the asymptotic expansion of the
matrix $\Phi(k)$ as $k\to\infty$ in the linear equation (\ref{RH3}). In
the $N$-wave interaction model it is given by formula (\ref{UV3wave}).
The pure soliton solutions are obtained by using the rational matrix
$\Phi=\Gamma(k)$.

The above set of discrete spectral data (\ref{vects}) holds only for the
generic case where zeros of $\det\Phi_+(k)$ and $\det\Phi_-^{-1}(k)$
are simple. If these zeros are higher-order rather than simple,
what the discrete spectral data should be and how they evolve with
$x$ and $t$ is still unknown yet. We have stressed in Sec. 1 that
the case of higher-order zeros can not be treated by coalescing simple
zeros, thus is highly non-trivial. In this paper, this problem
will be resolved completely.

\section{Soliton matrices for general higher-order zeros}
\label{multpl}

In this section we derive the soliton matrices for an arbitrary matrix
dimension $N$ and an arbitrary number of higher-order zeros under the
assumption that these zeros are normal (see Definition \ref{def1}).
Normal higher-order zeros are most common in practice.
In general, they are non-elementary.
Our approach is based on a generalization of
the idea in our previous paper \cite{SIAM}.

\subsection{Product representation of soliton matrices}
Our starting point to tackle this problem is to derive a product representation
for soliton matrices. This product representation is not convenient for
obtaining soliton solutions, but it will lead to the summation representation
of soliton matrices, which are very useful.

In treating the soliton matrix as a product of constituent matrices
(called elementary matrices in Ref.~\cite{NMPZ84}, see formulae (\ref{S4}) and (\ref{S6}) below)
one can consider each zero of the Riemann-Hilbert problem separately. For instance,
consider a pair of zeros $k_1$ and $\overline{k}_1$, respectively, of $\Phi_+(k)$
and $\Phi_-^{-1}(k)$ from Eq. (\ref{RH1}), each having order $m$:
\begin{equation} \label{det1}
\det\Phi_+(k) = (k-k_1)^m\varphi(k),\quad \det\Phi^{-1}_-(k) =
(k-\overline{k}_1)^m\overline{\varphi}(k),
\end{equation}
where  $\varphi(k_1)\ne0$ and $\overline{\varphi}(\overline{k}_1)\ne0$. The
geometric multiplicity of $k_1$ ($\overline{k}_1$) is defined as the number of
independent vectors in the kernel of $\Phi_+(k_1)$ ($\Phi_-^{-1}(\overline{k}_1)$),
see (\ref{vects}). In other words, the geometric multiplicity of $k_1$
($\overline{k}_1$) is the dimension of the kernel space of $\Phi_+(k_1)$
($\Phi_-^{-1}(\overline{k}_1)$). It can be easily shown that the order of a zero is
always greater or equal to its geometric multiplicity. It is also obvious that the
geometric multiplicity of a zero is less than the matrix dimension.   Let us recall
how the soliton matrices are usually constructed (see, for instance,
Refs.~\cite{NMPZ84,Kawata}). Starting from the solution $\Phi_\pm(k)$ to the
Riemann-Hilbert problem (\ref{RH1})-(\ref{RH2}), one looks for the independent
vectors in the kernels of the matrices $\Phi_+(k_1)$ and
$\Phi_-^{-1}(\overline{k}_1)$. Assuming that the geometric multiplicities of $k_1$
and $\overline{k}_1$ are the same and equal to $r_1$, then we have
\begin{equation}
\Phi_+(k_1)|v_{i1}\rangle = 0,\quad
\langle\overline{v}_{i1}|\Phi_{-}^{-1}(\overline{k}_1) = 0,\quad i = 1,\ldots,r_1.
\label{S3}\end{equation}
Next, one constructs the constituent matrix
\begin{equation}
\chi_1(k) = I -\frac{k_1-\overline{k}_1}{k-\overline{k}_1}P_1,\quad
\label{S4}\end{equation}
where
\begin{equation}
P_1 = \sum_{i, j}^{r_1}|v_{i1} \rangle(K^{-1})_{ij}\langle\overline{v}_{j 1}|,\quad
K_{ij} = \langle\overline{v}_{i1}|v_{j 1}\rangle.
\label{S5}\end{equation}
Here $P_1$ is a projector matrix, i.e., $P_1^2=P_1$. It can be shown that
$\mbox{det}\chi_1=(k-k_1)^{r_1}/(k-\overline{k}_1)^{r_1}$ [note that the geometric
multiplicity $r_1$ is equal to rank$P_1$]. If $r_1<m$ then one considers the new
matrix functions
\[\widetilde{\Phi}_+(k) =
\Phi_+(k)\chi_1^{-1}(k), \quad \quad \widetilde{\Phi}_-^{-1}(k)=\chi_1(k)\Phi_-^{-1}(k).\]
By virtue of equations (\ref{S3}), the matrices $\widetilde{\Phi}_+(k)$ and
$\widetilde{\Phi}_-^{-1}(k)$ are also holomorphic in the respective half planes of
the complex plane (see Lemma~1 in Ref.~\cite{SIAM}). In addition, $k_1$
and $\overline{k}_1$ are still zeros of $\mbox{det}\widetilde{\Phi}_+(k)$
and $\mbox{det}\widetilde{\Phi}_-^{-1}(k)$.
Assuming that the geometric multiplicities of zeros $k_1$ and $\overline{k}_1$
in new matrices $\widetilde{\Phi}_+(k)$
and $\widetilde{\Phi}_-^{-1}(k)$ are still the same and equal to $r_2$,
then the above steps can be repeated, and we can define matrix $\chi_2(k)$
analogous to Eq. (\ref{S4}).
In general, if the geometric multiplicities of zeros $k_1$ and $\overline{k}_1$
in matrices
\begin{equation} \label{submatrix}
\widetilde{\Phi}_+(k) =
\Phi_+(k)\chi_1^{-1}(k)\dots \chi_{l-1}^{-1}(k),
\quad \quad \widetilde{\Phi}_-^{-1}(k)=\chi_{l-1}(k)\dots\chi_1(k)\Phi_-^{-1}(k)
\end{equation}
are the same and given by $r_l$ ($l=1, 2, ...$), then we can define a matrix
$\chi_l$ similar to Eqs. (\ref{S4}) and (\ref{S5}) but the independent vectors
$|v_{il}\rangle$ and $\langle\overline{v}_{il}|$ ($i=1, \dots, r_l$) are from the
kernels of $\widetilde{\Phi}_+(k_1)$ and $\widetilde{\Phi}_-^{-1}(\overline{k}_1)$
in Eq. (\ref{submatrix}). When this process is finished, one would get the
constituent matrices $\chi_1(k)$, \ldots, $\chi_r(k)$ such that $r_1+r_2+\ldots +
r_n = m$, and the product representation of the soliton matrix $\Gamma(k)$,
\begin{equation}
\Gamma(k) = \chi_{n}(k)\cdots\chi_2(k)\chi_1(k),
\label{S6}\end{equation}

This product representation (\ref{S6}) is our starting point of this paper.
In arriving at this representation,
our assumptions are that the zeros $k_1$ and $\overline{k}_1$ have the same algebraic
multiplicity [see Eq. (\ref{det1})], and their geometric multiplicities
in matrices $\widetilde{\Phi}_+(k)$
and $\widetilde{\Phi}_-^{-1}(k)$ of Eq. (\ref{submatrix}) are also the same
for all $l$'s.
For convenience, we introduce the following definition.

\begin{defi}
\label{def1}
A pair of zeros $k_1$ and $\overline{k}_1$ in the matrix Riemann-Hilbert problem
are called normal zeros if they have the same algebraic multiplicity, and
their geometric multiplicities
in matrices $\widetilde{\Phi}_+(k)$
and $\widetilde{\Phi}_-^{-1}(k)$ of Eq. (\ref{submatrix}) are also the same
for all $l$'s.
\end{defi}

In the text of this paper, we only consider normal zeros of the matrix Riemann-Hilbert problem.
The case of abnormal zeros will be briefly discussed in the Appendix.

\noindent
{\bf Remark 1 } Under the involution property (\ref{invol}), all zeros are normal.
Thus, our results for normal zeros cover almost all the physically important
integrable PDEs.

\noindent
{\bf Remark 2 } Normal zeros include the elementary zeros of \cite{SIAM} as special cases, but
they are non-elementary in general.

\vspace{0.3cm}
It is an important fact (see Ref.~\cite{SIAM}, Lemma~2) that the
sequence of ranks of the projectors $P_l$ in the matrix
$\Gamma(k)$ given by Eq. (\ref{S6}), i.e. built in the
described way, is non-increasing:
\begin{equation}
\text{rank\,}P_n\le\text{rank\,}P_{r-1}\le\ldots\le\text{rank\,}P_1,
\label{S7}\end{equation}
i.e., $r_n\le r_{n-1}\le\ldots\le r_1$. This result allows one to classify all
possible occurrences of a higher-order zero of the Riemann-Hilbert problem for an
arbitrary matrix dimension $N$. In general, for zeros of the same order, different
sequences of ranks in Eq. (\ref{S7}) give different classes of higher-order soliton
solutions. In Ref.~\cite{SIAM} we constructed the soliton matrices for the simplest
sequence of ranks, i.e., 1,...,1. Such zeros are called ``elementary''. If the
matrix dimension $N=2$ (as for the nonlinear Schr\"odinger equation), then all
higher-order zeros are elementary since $\text{rank}P_1$ is always equal to 1.

To obtain the product representation for soliton matrices corresponding to several
higher-order normal zeros one can multiply the matrices of the type (\ref{S6}) for each
zero, i.e. $\Gamma(k) = \Gamma_1(k)\Gamma_2(k)\cdots\Gamma_{N_Z}(k)$,
where $N_Z$ is the number of distinct zeros and each $\Gamma_j(k)$ has the form
given by formula (\ref{S6}) with $n$ substituted by some $n_j$.

The product representation (\ref{S6}) of the soliton matrices is
difficult to use for actual calculations of the soliton solutions. Indeed, though
the representation (\ref{S6}) seems to be simple, derivation of the
$(x,t)$-dependence of the involved vectors (except for the vectors in the first
projector $P_1$) requires solving matrix equations with $(x,t)$-dependent
coefficients. One would like to have a more convenient representation, where
all the involved vectors have explicit $(x,t)$-dependence. Below we derive
such a representation for soliton matrices corresponding to an arbitrary number of
higher-order normal zeros.

For the sake of clarity, we consider first the case of a single pair
of higher-order zeros, followed by the most general case of several distinct pairs
of higher-order zeros.

\subsection{Soliton matrices for a single pair of zeros}
\label{3B}
Let us introduce a definition.

\begin{defi}
\label{def2}
For soliton matrices having a single pair of higher-order normal zeros
$(k_1,\overline{k}_1)$, suppose $\Gamma(k)$ is
constructed judiciously as in Eq. (\ref{S6}), with ranks $r_j$ of matrices $P_j
(1\le j\le n)$ satisfying inequality (\ref{S7}), i.e.,
\[ r_n \le r_{n-1} \le \dots \le r_1. \]
Then a new sequence of positive integers
\[s_1 \ge s_2 \ge \dots \ge s_{r_1} \]
are defined as follows:
\begin{center}
$s_\nu\equiv$ the index of the last positive integer in the array
$[r_1+1-\nu,r_2+1-\nu, \dots, r_{n}+1-\nu].$
\end{center}
We call the sequence of integers $\{r_n, r_{n-1}, \dots, r_1\}$ the rank sequence
associated with the pair of zeros $(k_1,\overline{k}_1)$, and the new sequence
$\{s_1, s_2, \dots, s_{r_1}\}$ the block sequence associated with this pair of zeros.
\end{defi}

\noindent{\bf Remark } It is easy to see that the sum of the block sequence is
equal to the sum of all ranks,
\[
\sum_{\nu=1}^{r_1}s_\nu = \sum_{l=1}^n r_l,
\]
with the latter being equal to the algebraic order of the Riemann-Hilbert zeros
$(k_1,\overline{k}_1)$.

For example, if the rank sequence is $\{3\}$ [only one constituent matrix in
(\ref{S6}) -- trivial higher-order zero], then the block sequence is $\{1,1,1\}$;
if the rank sequence is $\{1,1,1,1\}$ (an elementary zero), then the block sequence
is $\{4\}$; if the rank sequence is $\{2,3,5,7\}$, then the block sequence is
$\{4,4,3,2,2,1,1\}$.

With these definitions the most general soliton matrices $\Gamma(k)$ and
$\Gamma^{-1}(k)$ for a single pair of higher-order normal zeros $(k_1,\overline{k}_1)$ are
given as follows. This result is a generalization of our previous result \cite{SIAM} to
non-elementary higher-order zeros.

\begin{lem}
\label{lem1}
Consider a single pair of higher-order normal zeros $(k_1,\overline{k}_1)$ in the Riemann-Hilbert
problem. Suppose their geometric multiplicity
is $r_1$, and their block sequence is $\{s_1, s_2, \dots, s_{r_1}\}$.
Then the soliton matrices
$\Gamma(k)$ and $\Gamma^{-1}(k)$ can be written in the following summation forms:
\begin{equation}
\Gamma(k) = I + \sum_{\nu=1}^{r_1} \overline{{\cal S}}_\nu, \quad
\Gamma^{-1}(k) = I + \sum_{\nu=1}^{r_1} {\cal S}_\nu.
\label{S80}\end{equation}
Here ${\cal S}_\nu$ and $\overline{{\cal S}}_\nu$ are the following block matrices,
\begin{subequations}
\label{BLK0}
\begin{equation}
\overline{{\cal S}}_\nu =
\sum_{l=1}^{s_\nu}\sum_{j=1}^l\frac{|\overline{q}^{(\nu)}_j\rangle\langle
\overline{p}^{(\nu)}_{l+1-j}|} {(k-\overline{k}_1)^{s_\nu+1-l}} =
(|\overline{q}^{(\nu)}_{s_\nu}\rangle,\ldots,|\overline{q}^{(\nu)}_1\rangle)\overline{D}_{\nu
}(k)\left(\begin{array}{c} \langle \overline{p}^{(\nu)}_1| \\ \vdots\\ \langle
\overline{p}_{s_\nu}^{(\nu)}| \end{array}\right),
\end{equation}
\begin{equation}
{\cal S}_\nu = \sum_{l=1}^{s_\nu}\sum_{j=1}^l\frac{|p^{(\nu)}_{l+1-j}\rangle\langle
q^{(\nu)}_j|}{(k-{k}_1)^{s_\nu+1-l}} =
(|p^{(\nu)}_1\rangle,\ldots,|p^{(\nu)}_{s_\nu}\rangle)D_{\nu}(k)
\left(\begin{array}{c} \langle{q}^{(\nu)}_{s_\nu}| \\ \vdots \\ \langle
{q}^{(\nu)}_1|\end{array}\right),
\end{equation}
\end{subequations}
$D_{\nu}(k)$ and $\overline{D}_{\nu}(k)$ are the triangular Toeplitz matrices with poles:
\begin{equation}
\overline{D}_{\nu}(k) = \left(\begin{array}{cccc}
\frac{1}{(k-\overline{k}_{1})}&0&\ldots&0\\
\frac{1}{(k-\overline{k}_{1})^{2}}&\frac{1}{(k-\overline{k}_{1})}&\ddots&\vdots\\
\vdots&\ddots&\ddots&0\\
\frac{1}{(k-\overline{k}_{1})^{s_\nu}}& \ldots & \frac{1}{(k-\overline{k}_{1})^{2}}
&\frac{1}{(k-\overline{k}_{1})}\end{array}\right),\; D_{\nu}(k) =
\left(\begin{array}{cccc} \frac{1}{(k-k_{1})}&\frac{1}{(k-k_{1})^2}&\ldots&
\frac{1}{(k-k_{1})^{s_\nu}}\\
0&\ddots&\ddots&\vdots \\
\vdots&\ddots&\frac{1}{(k-k_{1})}&\frac{1}{(k-k_{1})^2}\\
0&\ldots&0&\frac{1}{(k-k_{1})}\end{array}\right).
\label{Dse}\end{equation}
The vectors $| p^{(\nu)}_i\rangle, \langle\overline{p}^{(\nu)}_i|, \langle
q^{(\nu)}_i|, |\overline{q}^{(\nu)}_i\rangle \; (i=1, \dots, s_\nu)$ here are
independent of $k$, and
each of the two sets of vectors \{$|p^{(1)}_1\rangle,...,|p^{(r_1)}_1\rangle$\} and
\{$\langle\overline{p}^{(1)}_1|,..., \langle\overline{p}^{(r_1)}_1|$\} are linearly
independent.
\end{lem}

\noindent{\bf Remark 1 } If $r_1=1$, the zeros $k_1$ and $\overline{k}_1$ are elementary
\cite{SIAM}. In this case, the above soliton matrices reduce to those in \cite{SIAM}.

\noindent{\bf Remark 2 } The total number of all $|p\rangle$-vectors or $\langle
\overline{p}|$-vectors from all $\nu$-blocks are equal to the algebraic order of the zeros
$k_1$ and $\overline{k}_1$.
\medskip

\noindent{\bf Proof} The representation (\ref{S80}) can be proved by induction.
Consider, for instance, the formula for $\Gamma(k)$. Obviously, this formula is
valid for $n=1$ in Eq. (\ref{S6}), where $\Gamma(k)$ contains only a single matrix
$\chi_1(k)$. Now, suppose that this formula  is valid for $n>1$. We need to show
that it is valid for $n+1$ as well. Indeed, denote the soliton matrices for $n$ and
$n+1$ by $\Gamma(k)$ and $\widetilde{\Gamma}(k)$ respectively, the rightmost
multiplier in $\widetilde{\Gamma}(k)$ being $\widetilde{\chi}(k)$. Then  we have
\[
\widetilde{\Gamma}(k) = \Gamma(k)\widetilde{\chi}(k) = \left(I
+\frac{{A}_1}{k-\overline{k}_1}+\frac{{A}_2}{(k-\overline{k}_1)^2}
+ \ldots+\frac{{A}_{n}}{(k-\overline{k}_1)^{n}}\right)\left( I +\frac{R}{k-\overline{k}_1}\right)
\]
\begin{equation}
=I +\frac{\widetilde{A}_1}{k-\overline{k}_1}+\frac{\widetilde{A}_2}{(k-\overline{k}_1)^2}
+ \ldots+\frac{\widetilde{A}_{n+1}}{(k-\overline{k}_1)^{n+1}},
\label{S9e}\end{equation}
where
\begin{equation}
R \equiv (\overline{k}_1-k_1)\widetilde{P} = \sum_{l=1}^{\tilde{r}}|u_l\rangle\langle \overline{u}_l|.
\label{Rdef}\end{equation}
Here we have normalized the
vectors $|u_l\rangle$ and $\langle\overline{u}_l|$ such that
\begin{equation} \label{unorm}
\langle\overline{u}_l|u_i\rangle =
(\overline{k}_1-k_1)\delta_{l,i},
\end{equation}
and $\tilde{r}=\mbox{rank}R$. In view of Eq. (\ref{S7}), we know that $\tilde{r}\ge
r_1$, where $r_1$ is the geometric multiplicity of $k_1$ and $\overline{k}_1$ in
the soliton matrices $\Gamma(k)$ and $\Gamma^{-1}(k)$. The coefficients at the
poles in $\widetilde{\Gamma}(k) $ are given by
\begin{equation}
\widetilde{A}_1 = A_1 + R,\quad\widetilde{A}_j = A_j + A_{j-1}R,\quad j =
2,...,n,\quad \widetilde{A}_{n+1} = A_{n}R.
\label{coeffA}\end{equation}
Consider first the coefficients $\widetilde{A}_2$ to $\widetilde{A}_{n+1}$.
The explicit form of coefficients $A_j$ can be obtained from Eqs. (\ref{S80}), (\ref{BLK0}),
and (\ref{S9e}) as
\begin{equation}
A_j \equiv \sum_{\nu=1}^{r_1}A_j^{(\nu)} =
\sum_{\nu=1}^{r_1}\sum_{l=1}^{s_\nu+1-j}|\overline{q}_l^{(\nu)}\rangle\langle
\overline{p}_{s_\nu+2-j-l}^{(\nu)}|,
\label{Aj}\end{equation}
where the inner sum is zero if $s_\nu+1-j\le0$. Substituting this expression into
(\ref{coeffA}) and defining the following new vectors in each block
\begin{equation}
\langle\widetilde{\overline{p}}_{1}^{(\nu)}| =
\langle\overline{p}^{(\nu)}_1|R,\quad \langle\widetilde{\overline{p}}_{j}^{(\nu)}|
= \langle\overline{p}^{(\nu)}_j|R+\langle\overline{p}^{(\nu)}_{j-1}|, \quad  j =
2,...,s_\nu,
\label{newps}\end{equation}
(for blocks of size 1, $s_\nu=1$, the second formula in (\ref{newps}) is dropped),
we then put the coefficients $\widetilde{A}_2,\ldots,\widetilde{A}_{n+1}$ into the
required form:
\[
\widetilde{A}_j=\sum_{\nu=1}^{r_1}\sum_{l=1}^{\tilde{s}_\nu+1-j}|\widetilde{\overline{q}}_l^{(\nu)}\rangle\langle
\widetilde{\overline{p}}_{\tilde{s}_\nu+2-j-l}^{(\nu)}|, \;\;\; j=2, \dots, n+1,
\]
where
\[|\widetilde{\overline{q}}_l^{(\nu)}\rangle \equiv |\overline{q}_l^{(\nu)}\rangle, \;\; l=1, \dots,
\tilde{s}_{\nu}-1, \]
and $\tilde{s}_\nu = s_\nu+1$, i.e., the size of
each $\nu$-block grows by one as we multiply by $\widetilde{\chi}(k)$ in formula
(\ref{S9e}).

Next, we consider the coefficient $\widetilde{A}_1$. Defining the vector
$\langle\widetilde{\overline{p}}_{\tilde{s}_\nu}^{(\nu)}| \equiv
\langle\overline{p}_{s_\nu}^{(\nu)}|$ and utilizing the definition
(\ref{newps}), we can rewrite $A_1^{(\nu)}$ as
\begin{equation}
{A}^{(\nu)}_1 =
\sum_{l=1}^{\tilde{s}_\nu-1}|\overline{q}^{(\nu)}_l\rangle\langle\widetilde{\overline{p}}^{(\nu)}_{\tilde{s}_\nu+1
-l}|-\sum_{l=2}^{s_\nu}|\overline{q}^{(\nu)}_l\rangle\langle{\overline{p}}^{(\nu)}_{s_\nu+2-l}|R.
\label{A1step}\end{equation}
To put $\widetilde{A}_1
= A_1+R$ into the required form
\begin{equation} \label{Atildeform}
\widetilde{A}_1=\sum_{\mu=r_1+1}^{\tilde{r}}|\widetilde{\overline{q}}_1^{(\mu)}\rangle\langle\widetilde{\overline{p}}^{(\mu)}_1|+
\sum_{\nu=1}^{r_1}\sum_{l=1}^{\tilde{s}_\nu}|\widetilde{\overline{q}}_l^{(\nu)}\rangle\langle
\widetilde{\overline{p}}_{\tilde{s}_\nu+1-l}^{(\nu)}|,
\end{equation}
we must define exactly one new vector
$|\widetilde{\overline{q}}_{\tilde{s}_\nu}^{(\nu)}\rangle$ for each $\nu$-block
[in the second term of Eq. (\ref{Atildeform})] and
$\tilde{r}-r_1$ new blocks of size 1 containing $2(\tilde{r}-r_1)$ new vectors
$|\widetilde{\overline{q}}_1^{(\mu)}\rangle$ and $\langle\widetilde{\overline{p}}^{(\mu)}_1|$.
Due to formulae (\ref{coeffA}) and
(\ref{A1step}), the new vectors to be defined must satisfy the following equation
\begin{equation}
\sum_{\mu=r_1+1}^{\tilde{r}}|\widetilde{\overline{q}}_1^{(\mu)}\rangle\langle\widetilde{\overline{p}}^{(\mu)}_1|
+\sum_{\nu=1}^{r_1}|\widetilde{\overline{q}}_{\tilde{s}_\nu}^{(\nu)}\rangle\langle\overline{p}^{(\nu)}_1|R
= R -
\sum_{\nu=1}^{r_1}\sum_{l=2}^{s_\nu}|\overline{q}_l^{(\nu)}\rangle\langle{\overline{p}}^{(\nu)}_{s_\nu+2-l}|R,
\label{tosolve}\end{equation}
where the $\langle \widetilde{\overline{p}}^{(\nu)}_1|$ definition in Eq. (\ref{newps})
has been utilized.
Substituting the expression (\ref{Rdef}) for $R$ into the above equation, we get
\begin{equation}\label{solve}
\sum_{\mu=r_1+1}^{\tilde{r}}|\widetilde{\overline{q}}_1^{(\mu)}\rangle\langle\widetilde{\overline{p}}^{(\mu)}_1|
=\sum_{l=1}^{\tilde{r}}|\xi_l\rangle \langle
\overline{u}_l |.
\end{equation}
where
\[|\xi_l\rangle \equiv
\left(I-
\sum_{\nu=1}^{r_1}\sum_{l=2}^{s_\nu}|\overline{q}_l^{(\nu)}\rangle\langle{\overline{p}}^{(\nu)}_{s_\nu+2-l}|\right)
|u_l\rangle-\sum_{\nu=1}^{r_1}|
\widetilde{\overline{q}}_{\tilde{s}_\nu}^{(\nu)}\rangle\langle\overline{p}^{(\nu)}_1|u_l\rangle,
\;\;\;\; l=1, \dots, \tilde{r}.
\]
To show that Eq. (\ref{solve}) is solvable, we need to use an important fact, i.e.,
the matrix
\[{\cal M}=({\cal M}_{\nu,l}), \;\;\; {\cal M}_{\nu,l}=
\langle{\overline{p}}^{(\nu)}_1|u_l\rangle, \quad
\nu=1, \dots, r_1, \;
l=1,...,\tilde{r}_1,
\]
has rank $r_1$. This fact can be proved by contradiction as follows.

Suppose the matrix ${\cal M}$ has rank less than $r_1$. Then
its $r_1$ rows are linearly dependent. Thus,
there are such scalars $C_1, C_2, \dots, C_{r_1}$, not equal to zero simultaneously,
that the vector
\[\langle \eta | \equiv \sum_{\nu=1}^{r_1} C_\nu \langle \overline{p}_1^{(\nu)}|\]
is orthogonal to all $|u_l\rangle$'s, i.e.,
\begin{equation} \label{eta}
\langle \eta | u_l \rangle =0, \quad  1\le l \le \tilde{r}.
\end{equation}
According to our induction assumption that soliton matrices for $n$ have the form
(\ref{S80}), we can easily show from the identity $\Gamma(k) \Gamma^{-1}(k)=I$ that
$\langle \overline{p}_1^{(\nu)}| \Gamma^{-1}(\overline{k}_1)$=0 for all $1\le \nu
\le r_1$ (see \cite{SIAM}). Thus $\langle \eta | \Gamma^{-1}(\overline{k}_1)=0$ as
well. According to Lemma 1 in \cite{SIAM}, if $\langle \eta|$ is in the kernel of
$\Gamma^{-1}(\overline{k}_1)$ and is orthogonal to all $|u_l\rangle$'s, then $\langle
\eta|$ is in the kernel of $\widetilde{\Gamma}^{-1}(\overline{k}_1)$ as well, i.e.,
$\langle \eta| \widetilde{\Gamma}^{-1}(\overline{k}_1)=0$. But according to our
construction of soliton matrices [see Eq. (\ref{S6})], the vectors
$\langle\overline{u}_l|$ $(l = 1,...,\tilde{r})$ are all the linearly independent
vectors in the kernel of $\widetilde{\Gamma}^{-1}(\overline{k}_1)$. Thus $\langle
\eta|$ must be a linear combination of $\langle\overline{u}_l|$'s. Then in view of Eqs.
(\ref{unorm}) and (\ref{eta}), we find that $\langle \eta|=0$, which leads to a
contradiction.

Now that the matrix ${\cal M}$ has rank $r_1$, then we are able to select vectors
$|\widetilde{\overline{q}}_{\tilde{s}_\nu}^{(\nu)}\rangle \; (\nu=1, \dots, r_1)$ such that
$r_1$ of the $\tilde{r}$ vectors $\langle \xi_l|$ are zero. With this choice of
$|\widetilde{\overline{q}}_{\tilde{s}_\nu}^{(\nu)}\rangle$'s, the r.h.s. of Eq. (\ref{solve})
becomes $\tilde{r}-r_1$ blocks of size 1. Assigning these blocks to the l.h.s. of (\ref{solve}),
then Eq. (\ref{solve}) can be solved. Hence
we can put the coefficient $\widetilde{A}_1$ in the required form
(\ref{Atildeform}).

Next we prove that all vectors $\langle
\widetilde{\overline{p}}^{(\nu)}_1|$ $(1\le \nu \le \tilde{r})$
in the matrix $\widetilde{\Gamma}(k)$
are linearly independent.
These vectors were defined in the above proof as
\begin{equation} \label{newp1}
\langle\widetilde{\overline{p}}_{1}^{(\nu)}| =
\langle\overline{p}^{(\nu)}_1|R=\sum_{l=1}^{\tilde{r}}\langle \overline{p}_1^{(\nu)}|u_l\rangle
\langle \overline{u}_l|, \quad 1\le \nu \le r_1,
\end{equation}
and $\langle\widetilde{\overline{p}}_{1}^{(\nu)}|$ for $r_1+1 \le \nu\le \tilde{r}$
are simply equal to $\tilde{r}-r_1$ of the vectors $\overline{u}_l$ depending on
what $r_1\times r_1$ submatrix of ${\cal M}$ has rank $r_1$.
To be definite, let us suppose the first $r_1$ columns of the matrix ${\cal M}$
have rank $r_1$ (i.e., linearly independent). Then according to the above proof,
we can uniquely select vectors
$|\widetilde{\overline{q}}_{\tilde{s}_\nu}^{(\nu)}\rangle \; (\nu=1, \dots, r_1)$
such that $|\xi_l\rangle=0$ for $1\le l\le r_1$. Thus,
\begin{equation} \label{newp2}
\langle\widetilde{\overline{p}}^{(\nu)}_1|=\langle \overline{u}_{\nu}|, \quad
r_1+1 \le \nu \le \tilde{r}.
\end{equation}
Recalling that vectors $\langle \overline{u}_\nu|$ $(1\le \nu \le \tilde{r})$ in the
projector $R$ (\ref{Rdef}) are linearly independent, and
the first $r_1$ columns of matrix ${\cal M}$ have rank $r_1$, we easily see that
vectors $\langle\widetilde{\overline{p}}_{1}^{(\nu)}|$ $(1\le \nu \le \tilde{r})$
as defined in Eqs. (\ref{newp1}) and (\ref{newp2}) are linearly independent.

Lastly, we prove that the sizes of blocks in representations (\ref{S80})
are given by the block sequence defined in Definition~\ref{def2}.
An equivalent statement is that the numbers of matrix blocks with
sizes $[1,2,3,...,n]$ are given by the pair-wise differences in the sequence of
ranks: $[r_1-r_2, r_2-r_3, ..., r_{n-1}-r_n,r_n]$, where the last number in the
sequence defines the number of blocks of size $n$. This can be
easily proven by the induction argument using the fact that the number of new
blocks of size 1 in $\widetilde{A}_1$ (\ref{coeffA}) is given by $\tilde{r}-r_1$,
while the sizes of old blocks grow by 1 in each multiplication as in formula
(\ref{S9e}).

Using similar arguments, we can prove that the representation (\ref{S80}) for
$\Gamma^{-1}(k)$ is valid, and vectors
$|p^{(1)}_1\rangle,...,|p^{(r_1)}_1\rangle$ are linearly
independent. This concludes the proof of Lemma \ref{lem1}. Q.E.D.

\subsection{Soliton matrices for several pairs of zeros}
Next, we extend the above results to the most general case of several pairs of
higher-order normal zeros $\{(k_1,\overline{k}_1),\ldots,(k_{N_Z},\overline{k}_{N_Z})\}$.
In this general case, the soliton matrix $\Gamma(k)$ can be constructed as
a product of soliton matrices (\ref{S6}) for each zero by the procedure
layed out in the beginning of this section [see Eqs. (\ref{det1}) to (\ref{S6})].
Thus, $\Gamma(k)$ can be represented as
\begin{equation}
\Gamma(k) = \Gamma_1(k)\cdot\Gamma_2(k)\cdots\Gamma_{N_Z}(k).
\label{gengam}\end{equation}
For each pair of zeros $(k_n, \overline{k}_n)$, we can define its rank sequence and block sequence
by Definition \ref{def2} either from $\Gamma(k)$ directly or from
the individual matrix $\Gamma_n(k)$ associated with this zero.
It is easy to see that using $\Gamma(k)$ or $\Gamma_n(k)$ gives identical results.
The inverse matrix $\Gamma^{-1}(k)$ can be represented in a similar way.

The product representation (\ref{gengam}) for $\Gamma(k)$ and its counterpart for
$\Gamma^{-1}(k)$ are not convenient for deriving soliton solutions.
Their summation representations such as Eq. (\ref{S80}) are needed.
It turns out that $\Gamma(k)$ and $\Gamma^{-1}(k)$ in the general case
are given simply by sums of all
the blocks from all pairs of zeros plus the unit matrix. Let us formulate this
result in the next lemma.

\begin{lem}
\label{lem2}
Consider several pairs of higher-order normal zeros
$\{(k_1,\overline{k}_1),\ldots,(k_{N_Z},\overline{k}_{N_Z})\}$ in the Riemann-Hilbert
problem. Denote
the geometric multiplicity of zeros $(k_n,\overline{k}_n)$ as
$r_1^{(n)}$, and their block sequence as
$\{s_1^{(n)}, s_2^{(n)}, \dots, s_{r_1^{(n)}}^{(n)}\}$ ($1\le n \le N_Z$).
Then the soliton matrices
$\Gamma(k)$ and $\Gamma^{-1}(k)$ can be written in the following summation forms:
\begin{equation}
\Gamma(k) = I + \sum_{n=1}^{N_Z}\sum_{\nu=1}^{r_1^{(n)}} \overline{{\cal S}}_\nu^{(n)}, \quad
\Gamma^{-1}(k) = I + \sum_{n=1}^{N_Z}\sum_{\nu=1}^{r_1^{(n)}} {\cal S}_\nu^{(n)}.
\label{S8}
\end{equation}
Here ${\cal S}_\nu^{(n)}$ and $\overline{{\cal S}}_\nu^{(n)}$ are
the following block matrices,
\begin{subequations}
\label{BLK}
\begin{equation}
\overline{{\cal S}}_\nu^{(n)} =
(|\overline{q}_{s_\nu^{(n)}}^{(\nu,n)}\rangle,\ldots,|\overline{q}_{1}^{(\nu,n)}\rangle)\overline{D}_{\nu
}^{(n)}(k)\left(\begin{array}{c} \langle \overline{p}_{1}^{(\nu,n)}| \\ \vdots\\ \langle
\overline{p}_{s_\nu^{(n)}}^{(\nu,n)}| \end{array}\right),
\end{equation}
\begin{equation}
{\cal S}_\nu^{(n)} =
(|p_{1}^{(\nu,n)}\rangle,\ldots,|p_{s_\nu^{(n)}}^{(\nu,n)}\rangle)D_{\nu}^{(n)}(k)
\left(\begin{array}{c} \langle{q}_{s_\nu^{(n)}}^{(\nu,n)}| \\ \vdots \\ \langle
{q}_{1}^{(\nu,n)}|\end{array}\right),
\end{equation}
\end{subequations}
$D_{\nu}^{(n)}(k)$ and
$\overline{D}_{\nu}^{(n)}(k)$ are the triangular Toeplitz matrices with poles:
\begin{equation}
\overline{D}_{\nu}^{(n)}(k) = \left(\begin{array}{cccc}
\frac{1}{(k-\overline{k}_{n})}&0&\ldots&0\\
\frac{1}{(k-\overline{k}_{n})^{2}}&\frac{1}{(k-\overline{k}_{n})}&\ddots&\vdots\\
\vdots&\ddots&\ddots&0\\
\frac{1}{(k-\overline{k}_{n})^{s_\nu^{(n)}}}& \ldots &
\frac{1}{(k-\overline{k}_{n})^{2}}
&\frac{1}{(k-\overline{k}_{n})}\end{array}\right),\; D_{\nu}^{(n)}(k) =
\left(\begin{array}{cccc} \frac{1}{(k-k_{n})}&\frac{1}{(k-k_{n})^2}&\ldots&
\frac{1}{(k-k_{n})^{s_\nu^{(n)}}}\\
0&\ddots&\ddots&\vdots \\
\vdots&\ddots&\frac{1}{(k-k_{n})}&\frac{1}{(k-k_{n})^2}\\
0&\ldots&0&\frac{1}{(k-k_{n})}\end{array}\right).
\label{Ds}\end{equation}
Vectors $|p_{i}^{(\nu,n)}\rangle, \langle\overline{p}_{i}^{(\nu,n)}|, \langle
q_{i}^{(\nu,n)}|, |\overline{q}_{i}^{(\nu,n)}\rangle \; (i=1, \dots, s_\nu^{(n)})$ are
independent of $k$. In addition, for each $n$, vectors
\{$|p^{(1, n)}_1\rangle,...,|p^{(r_1^{(n)},n)}_1\rangle$\} and
\{$\langle\overline{p}^{(1,n)}_1|,..., \langle\overline{p}^{(r_1^{(n)},n)}_1|$\}
are linearly independent respectively.
\end{lem}

\noindent{\bf Remark} When there is only a single pair of zeros $(k_1,
\overline{k}_1)$, the above lemma reduces to Lemma \ref{lem1}.

\vspace{0.4cm} \noindent{\bf Proof} Again we will rely on the induction argument.
As it was already mentioned, the general soliton matrix $\Gamma(k)$ corresponding
to several distinct zeros can be represented as a product (\ref{gengam}) of
individual soliton matrices (\ref{S6}) for each zero. For clarity reason and
simplicity of the presentation we will give detailed calculations for the simplest
case of just one product in (\ref{gengam}). Then we will show how to generalize the
calculations. Consider soliton matrix $\Gamma(k)$ for two pairs of distinct
higher-order zeros $(k_1,\overline{k}_1)$ and $(k_2,\overline{k}_2)$. We have
$\Gamma(k) = \Gamma_1(k)\Gamma_2(k)$ and
\begin{equation}
\Gamma(k) = \left(I +\frac{{A}_1}{k-\overline{k}_1} +
\ldots+\frac{{A}_{n_1}}{(k-\overline{k}_1)^{n_1}}\right)\left(I
+\frac{{B}_1}{k-\overline{k}_2}+
\ldots+\frac{{B}_{n_2}}{(k-\overline{k}_2)^{n_2}}\right).
\label{gamAB}\end{equation}
Here $n_j\; (j=1, 2)$ is the number of simple matrices in the product representation (\ref{S6}) for
$\Gamma_j$. Due to Lemma \ref{lem1},
the coefficients $A_j$ and $B_j$ are given by formulae similar to (\ref{Aj}):
\begin{equation}
A_j =\sum_{\nu=1}^{r_1^{(1)}}\sum_{l=1}^{s_\nu^{(1)}+1-j}|\overline{q}_l^{(\nu, 1)}\rangle\langle
\overline{p}_{s_\nu^{(1)}+2-j-l}^{(\nu,1)}|,
\label{Ajnew}
\end{equation}
\begin{equation}
B_j =\sum_{\nu=1}^{r_1^{(2)}}\sum_{l=1}^{s_\nu^{(2)}+1-j}|\overline{q}_l^{(\nu, 2)}\rangle\langle
\overline{p}_{s_\nu^{(2)}+2-j-l}^{(\nu,2)}|.
\label{Bjnew}
\end{equation}
On the other hand, by expanding formula (\ref{gamAB}) into the partial fractions
we get
\begin{equation}
\Gamma(k) = I +\frac{\widetilde{A}_1}{k-\overline{k}_1}+
\ldots+\frac{\widetilde{A}_{n_1}}{(k-\overline{k}_1)^{n_1}}+
\frac{\widetilde{B}_1}{k-\overline{k}_2}+
\ldots+\frac{\widetilde{B}_{n}}{(k-\overline{k}_2)^{n_2}}.
\label{gamA+B}\end{equation}
Consider first the coefficients $\widetilde{A}_j$. Multiplication by
$(k-\overline{k}_1)^{n_1}$ of both formulae (\ref{gamAB}) and (\ref{gamA+B}) and
taking derivatives at $k=\overline{k}_1$  using the Leibniz rule gives
\begin{equation}
\widetilde{A}_{n_1-l} =\frac{1}{l!}\left\{\frac{\mathrm{d}^l}{\mathrm{d}k
{}^l}(k-\overline{k}_1)^{n_1}\Gamma(k)\right\}_{k=\overline{k}_1} =\sum_{j=0}^l
\frac{A_{n_1-j}}{(l-j)!}\frac{\mathrm{d}^{(l-j)}\Gamma_2}{\mathrm{d}k
{}^{(l-j)}}(\overline{k}_1).
\label{Anew}\end{equation}
In similar way we get
\begin{equation}
\widetilde{B}_{n_2-l}= \sum_{j=0}^l\frac{\mathrm{d}^{(l-j)}\Gamma_1}{\mathrm{d}k
{}^{(l-j)}}(\overline{k}_2)\frac{B_{n_2-j}}{(l-j)!}.
\label{Bnew}\end{equation}
Now substituting Eqs. (\ref{Ajnew}) and (\ref{Bjnew}) into (\ref{Anew}) and (\ref{Bnew})
and defining new vectors
\begin{equation}
\langle\widetilde{\overline{p}}^{(\nu,1)}_m| =
\sum_{j=0}^{m-1}\langle\overline{p}^{(\nu,1)}_{m-j}|\frac{1}{j!}\frac{\mathrm{d}^j\Gamma_2}{\mathrm{d}k
{}^j}(\overline{k}_1),\quad m=1,\ldots,s_\nu^{(1)},
\label{vectinA}\end{equation}
and
\begin{equation}
|\widetilde{\overline{q}}^{(\nu,2)}_m\rangle =
\sum_{j=0}^{m-1}\frac{1}{j!}\frac{\mathrm{d}^j\Gamma_1}{\mathrm{d}k
{}^j}(\overline{k}_2)|\overline{q}^{(\nu,2)}_{m-j}\rangle,\quad m=1,\ldots,s_\nu^{(2)},
\label{vectinB}\end{equation}
we find that
\begin{equation}
\widetilde{A}_j =\sum_{\nu=1}^{r_1^{(1)}}\sum_{l=1}^{s_\nu^{(1)}+1-j}|\overline{q}_l^{(\nu, 1)}\rangle\langle
\widetilde{\overline{p}}_{s_\nu^{(1)}+2-j-l}^{(\nu,1)}|,
\end{equation}
\begin{equation}
\widetilde{B}_j =\sum_{\nu=1}^{r_1^{(2)}}\sum_{l=1}^{s_\nu^{(2)}+1-j}|\widetilde{\overline{q}}_l^{(\nu, 2)}\rangle\langle
\overline{p}_{s_\nu^{(2)}+2-j-l}^{(\nu,2)}|,
\end{equation}
which give precisely the needed representation (\ref{S8}).
Note from definitions (\ref{vectinA}) and (\ref{vectinB}) that
\[
\left[\langle\widetilde{\overline{p}}^{(\nu,1)}_1|, \dots,
\langle\widetilde{\overline{p}}^{(\nu,1)}_{r_1^{(1)}}| \right]=
\left[\langle \overline{p}^{(\nu,1)}_1|, \dots,
\langle \overline{p}^{(\nu,1)}_{r_1^{(1)}}| \right] \Gamma_2(\overline{k}_1),
\]
and
\[
\left[|\widetilde{\overline{q}}^{(\nu,2)}_1\rangle, \dots,
|\widetilde{\overline{q}}^{(\nu,2)}_{r_1^{(2)}}\rangle\right]
=\Gamma_1(\overline{k}_2) \left[|\overline{q}^{(\nu,2)}_1\rangle, \dots,
|\overline{q}^{(\nu,2)}_{r_1^{(2)}}\rangle\right].
\]
Due to lemma \ref{lem1},
vectors $\{\langle \overline{p}^{(\nu,1)}_1|, \dots,
\langle \overline{p}^{(\nu,1)}_{r_1^{(1)}}|\}$
and $\{|\overline{q}^{(\nu,2)}_1\rangle, \dots,
|\overline{q}^{(\nu,2)}_{r_1^{(2)}}\rangle\}$ are linearly independent respectively.
In addition, matrices $\Gamma_1(\overline{k}_2)$ and $\Gamma_2(\overline{k}_1)$
are non-degenerate. Thus new vectors
$\{\langle\widetilde{\overline{p}}^{(\nu,1)}_1|, \dots,
\langle\widetilde{\overline{p}}^{(\nu,1)}_{r_1^{(1)}}|\}$
and
$\{|\widetilde{\overline{q}}^{(\nu,2)}_1\rangle, \dots,
|\widetilde{\overline{q}}^{(\nu,2)}_{r_1^{(2)}}\rangle\}$ are linearly independent
respectively as well. This completes the proof of Lemma \ref{lem2} for
two pairs of higher-order zeros.

It is easy to see that the above procedure of redefining the vectors in the blocks
corresponding to different zeros will also work in the general case, when $\Gamma_1(k)$
is replaced by the product $\Gamma_1(k)\cdot\dots\cdot\Gamma_n(k)$, and
$\Gamma_2(k)$ replaced by $\Gamma_{n+1}(k)$.
In this case, the sum over all distinct poles will
be present in the left $( )$-bracket in formula (\ref{gamAB}), and consequently
there will be more terms in formula (\ref{gamA+B}).
Formula (\ref{Anew}) will be
valid for coefficients $\widetilde{A}$ of each zero, and
formula (\ref{Bnew}) remains valid as well. Thus
by defining vectors
$\langle\widetilde{\overline{p}}^{(\nu,j)}_m|$ by formula (\ref{vectinA})
for each zero $k_j \; (1\le j\le n)$, and defining vectors
$|\widetilde{\overline{q}}^{(\nu,n+1)}_m\rangle$ by formula (\ref{vectinB}) for
zero $k_{n+1}$, we can show that the matrix $\Gamma(k)$ consisting of $n+1$ products
of $\Gamma_j(k)$ can be put in the required form (\ref{S8}).
This induction argument then completes the proof of Lemma \ref{lem2}. Q.E.D.

\vspace{0.5cm} The notations in the representation (\ref{S8}) for soliton matrices
with several zeros are getting complicated. To facilitate the presentations of
results in the remainder of this paper, let us reformulate the representation
(\ref{S8}). For this purpose, we define $r_1 = r_1^{(1)}+\ldots+r_1^{(N_Z)}$, where
$r_1^{(n)}$'s are as given in Lemma \ref{lem2}. Then we replace the double
summations in Eq. (\ref{S8}) with single ones,
\begin{equation}
\Gamma(k) = I + \sum_{\nu=1}^{r_1} \overline{{\cal S}}_\nu, \quad \Gamma^{-1}(k) =
I + \sum_{\nu=1}^{r_1} {\cal S}_\nu.
\label{S82}\end{equation}
Inside these single summations, the first $r_1^{(1)}$ terms are
blocks of type (\ref{BLK}) for the first pair of zeros $(k_1, \overline{k}_1)$,
the next $r_1^{(2)}$ terms are
blocks of type (\ref{BLK}) for the second pair of zeros $(k_2, \overline{k}_2)$,
and so on.
Block matrices ${\cal S}_\nu$ and $\overline{{\cal S}}_\nu$ can be written as
\begin{subequations}
\label{BLK2}
\begin{equation}
\overline{{\cal S}}_\nu =
\sum_{l=1}^{s_\nu}\sum_{j=1}^l\frac{|\overline{q}^{(\nu)}_j\rangle\langle
\overline{p}^{(\nu)}_{l+1-j}|} {(k-\overline{\kappa}_\nu)^{s_\nu+1-l}} =
(|\overline{q}^{(\nu)}_{s_\nu}\rangle,\ldots,|\overline{q}^{(\nu)}_1\rangle)\overline{D}_{\nu
}(k)\left(\begin{array}{c} \langle \overline{p}^{(\nu)}_1| \\ \vdots\\ \langle
\overline{p}_{s_\nu}^{(\nu)}| \end{array}\right),
\end{equation}
\begin{equation}
{\cal S}_\nu = \sum_{l=1}^{s_\nu}\sum_{j=1}^l\frac{|p^{(\nu)}_{l+1-j}\rangle\langle
q^{(\nu)}_j|}{(k-{\kappa}_\nu)^{s_\nu+1-l}} =
(|p^{(\nu)}_1\rangle,\ldots,|p^{(\nu)}_{s_\nu}\rangle)D_{\nu}(k)
\left(\begin{array}{c} \langle{q}^{(\nu)}_{s_\nu}| \\ \vdots \\ \langle
{q}^{(\nu)}_1|\end{array}\right),
\end{equation}
\end{subequations}
where matrices $D_{\nu}(k)$ and
$\overline{D}_{\nu}(k)$ are triangular Toeplitz matrices with poles:
\begin{equation}
\overline{D}_{\nu}(k) = \left(\begin{array}{cccc}
\frac{1}{(k-\overline{\kappa}_{\nu})}&0&\ldots&0\\
\frac{1}{(k-\overline{\kappa}_{\nu})^{2}}&\frac{1}{(k-\overline{\kappa}_{\nu})}&\ddots&\vdots\\
\vdots&\ddots&\ddots&0\\
\frac{1}{(k-\overline{\kappa}_{\nu})^{s_\nu}}& \ldots &
\frac{1}{(k-\overline{\kappa}_{\nu})^{2}}
&\frac{1}{(k-\overline{\kappa}_{\nu})}\end{array}\right),\; D_{\nu}(k) =
\left(\begin{array}{cccc} \frac{1}{(k-\kappa_{\nu})}&\frac{1}{(k-\kappa_{\nu})^2}&\ldots&
\frac{1}{(k-\kappa_{\nu})^{s_\nu}}\\
0&\ddots&\ddots&\vdots \\
\vdots&\ddots&\frac{1}{(k-\kappa_{\nu})}&\frac{1}{(k-\kappa_{\nu})^2}\\
0&\ldots&0&\frac{1}{(k-\kappa_{\nu})}\end{array}\right).
\label{Ds2}\end{equation}
Here
\begin{equation} \label{kappa}
\kappa_\nu=k_j,  \quad \mbox{if} \;\; 1+\sum_{l=1}^{j-1}r_1^{(l)}
\le \nu\le \sum_{l=1}^{j}r_1^{(l)}  \quad (1\le j \le N_Z).
\end{equation}
In other words, $\kappa_\nu=k_1$ for $1\le \nu\le r_1^{(1)}$,
$\kappa_\nu=k_2$ for $r_1^{(1)}+1\le \nu \le r_1^{(1)}+r_1^{(2)}$, etc.
In addition,
$\{s_\nu, 1+\sum_{l=1}^{j-1}r_1^{(l)}
\le \nu\le \sum_{l=1}^{j}r_1^{(l)}\}$ is the block sequence
of the $j$-th pair of zeros $(k_j, \overline{k}_j)$.
This new representation (\ref{S82}) is equivalent to (\ref{S8}), but it proves to be
helpful in the calculations below.

We note that the economical way of block numeration used in the representation (\ref{S82})
reflects the important property of the solitons matrices:  the soliton matrices
preserve their form if some of the zeros coalesce (or, vise versa, a zero splits
itself into two or more  zeros). The only thing that does change is the association
of a particular $\nu$-block to the pair of zeros.

\vspace{0.3cm}
The representation (\ref{S82}) [or (\ref{S8})] is but the first step towards the
necessary formulae for the soliton matrices. Indeed, there are twice as many
vectors in the expressions (\ref{S82}) for $\Gamma(k)$ and $\Gamma^{-1}(k)$ as
compared to the total number of vectors in the constituent matrices in the product
of representations of the type (\ref{S6}) for each pair of zeros. As the result,
only half of the vector parameters, say $|p^{(\nu)}_i\rangle$ and
$\langle\overline{p}^{(\nu)}_i|$, are free. To derive the formulae for the rest of
the vector parameters in (\ref{S82}) we can use the identity
$\Gamma(k)\Gamma^{-1}(k) = \Gamma^{-1}(k)\Gamma(k)= I$. First of all, let us give
the equations for the free vectors themselves.

\begin{lem}
\label{lem3}
The vectors $|p^{(\nu)}_1\rangle,\ldots,|p^{(\nu)}_{s_\nu}\rangle$ and
$\langle\overline{p}^{(\nu)}_1|,\ldots,\langle\overline{p}^{(\nu)}_{s_\nu}|$ from
each $\nu$-th block in the representation (\ref{S82})-(\ref{BLK2}) satisfy the
following linear systems of equations:
\begin{equation}
{\bf\Gamma}_{\nu}(\kappa_\nu)\left(\begin{array}{c} |p^{(\nu)}_1\rangle \\ \vdots \\
|p^{(\nu)}_{s_\nu}\rangle
\end{array}\right) = 0, \qquad
{\bf\Gamma}_{\nu}(k) \equiv \left(\begin{array}{cccc}
\Gamma&0&\ldots&\quad 0\\
\frac{1}{1!}\frac{\mathrm{d}}{\mathrm{d}k}\Gamma&\Gamma&\ddots&\quad\vdots\\
\vdots&\ddots&\ddots&\quad 0\\
\frac{1}{(s_\nu-1)!}\frac{\mathrm{d}^{s_\nu-1}}{\mathrm{d}k^{s_\nu-1}}\Gamma&
\ldots & \frac{1}{1!}\frac{\mathrm{d}}{\mathrm{d}k}\Gamma &\quad\Gamma
\end{array}\quad\right),
\label{S14}\end{equation}
\begin{equation}
\left(\langle\overline{p}^{(\nu)}_1|,\ldots,\langle\overline{p}^{(\nu)}_{s_\nu}|\right)
\overline{\bf\Gamma}_{\nu}(\overline{\kappa}_\nu) = 0,\qquad \overline{\bf\Gamma}_{\nu}(k)
\equiv \left(\begin{array}{cccc} \Gamma^{-1}\quad
&\frac{1}{1!}\frac{\mathrm{d}}{\mathrm{d}k}\Gamma^{-1}\quad &\ldots\quad &
\frac{1}{(s_\nu-1)!}\frac{\mathrm{d}^{s_\nu-1}}{\mathrm{d}k^{s_\nu-1}}\Gamma^{-1}\\
0\quad &\Gamma^{-1}\quad &\ddots\quad & \vdots  \\
\vdots\quad &\ddots\quad &\ddots\quad &
\frac{1}{1!}\frac{\mathrm{d}}{\mathrm{d}k}\Gamma^{-1} \\
0\quad &\ldots \quad &0 \quad &\Gamma^{-1}
\end{array}\right).
\label{S15}\end{equation}
\end{lem}

\noindent{\bf Remark } Note that the matrices ${\bf\Gamma}_{\nu}(k)$ and
${\bf\Gamma}^{-1}_\nu(k)$ have block-triangular Toeplitz forms, i.e., they have the
same (matrix) element along each diagonal.
\medskip

\noindent{\bf Proof} The derivation of the systems (\ref{S14})-(\ref{S15}) exactly
reproduces the analogous derivation in Ref.~\cite{SIAM} for the case of elementary
zeros (as the equations for the $\nu$-th block resemble analogous equations for a
single block corresponding to a pair of elementary zeros). For instance, the system
(\ref{S14}) is derived by considering the poles of $\Gamma(k)\Gamma^{-1}(k)$ at
$k=\kappa_\nu$, starting from the highest pole and using the representation
(\ref{S82})-(\ref{BLK2}) for $\Gamma^{-1}(k)$. The details are trivial and will not
be reproduced here.  Note that there may be several sets of vectors (from different
$\nu$-blocks of the same pair of zeros) which satisfy similar equations if the
geometric multiplicity of this pair of zeros is higher than 1. Q.E.D.

Now let us express the $|\overline{q}\rangle$- and $\langle q|$-vectors in the
expressions (\ref{S82})-(\ref{BLK2}) for $\Gamma(k)$ and $\Gamma^{-1}(k)$  through
the $|p\rangle$- and $\langle\overline{p}|$-vectors. This will lead to the needed
representation of the soliton matrices given through the $|p\rangle$- and
$\langle\overline{p}|$-vectors only. It is convenient to formulate the result in
the following lemma.

\begin{lem}
\label{lem4}
The general soliton matrices for several pairs of normal zeros
$\{(k_1,\overline{k}_1),\ldots,(k_{N_Z},\overline{k}_{N_Z})\}$
are given by the following formulae:
\begin{subequations}
\label{mainrepr}
\begin{equation}
\Gamma(k) = I -
(|p^{(1)}_1\rangle,\ldots,|p^{(1)}_{s_1}\rangle,\ldots,|p^{(r_1)}_1\rangle,\ldots,|p^{(r_1)}_{s_{r_1}}\rangle)
\overline{\cal K}{}^{-1}\overline{\cal D}(k)\left(\begin{array}{c}
\langle\overline{p}^{(1)}_1|\\ \vdots\\ \langle\overline{p}^{(1)}_{s_1}| \\
\vdots \\ \langle\overline{p}^{(r_1)}_1| \\ \vdots \\
\langle\overline{p}^{(r_1)}_{s_{r_1}}|\end{array}\right),
\label{mainrepr1}\end{equation}
\begin{equation}
\Gamma^{-1}(k) = I -
(|p^{(1)}_1\rangle,\ldots,|p^{(1)}_{s_1}\rangle,\ldots,|p^{(r_1)}_1\rangle,\ldots,|p^{(r_1)}_{s_{r_1}}\rangle)
{\cal D}(k){\cal K}^{-1}\left(\begin{array}{c}
\langle\overline{p}^{(1)}_1|\\ \vdots\\ \langle\overline{p}^{(1)}_{s_1}| \\
\vdots \\ \langle\overline{p}^{(r_1)}_1| \\ \vdots \\
\langle\overline{p}^{(r_1)}_{s_{r_1}}|\end{array}\right),
\label{mainrepr2}\end{equation}
\end{subequations}
where $s_\nu$ and $r_1$ are the same as in Lemma~\ref{lem2}. The matrices
$\overline{\cal D}(k)$ and ${\cal D}(k)$ are block-diagonal:
\begin{equation}
\overline{\cal D}(k) \equiv \left(\begin{array}{ccc} \overline{D}_1(k)  & \null & 0\\
\null & \ddots & \null \\ 0 & \null  & \overline{D}_{r_1}(k)
\end{array}\right),\quad
{\cal D}(k) \equiv \left(\begin{array}{ccc}{D}_1(k) & \null & 0\\
\null & \ddots & \null \\ 0 & \null & {D}_{r_1}(k) \end{array}\right),
\label{DDbar}\end{equation}
where the triangular Toeplitz matrices $\overline{D}_\nu(k)$ and ${D}_\nu(k)$ are
defined in formulae (\ref{Ds2}). The matrices $\overline{K}$ and $K$  have the
following block matrix representation:
\begin{equation}
\overline{\cal K} \equiv \left(\begin{array}{ccc} \overline{K}^{(1,1)} & \ldots &
\overline{K}^{(1,r_1)}\\ \vdots & \null & \vdots\quad \\
\overline{K}^{(r_1,1)} & \ldots & \overline{K}^{(r_1,r_1)}\end{array}\right),\quad
{\cal K} \equiv\left(\begin{array}{ccc}  {K}^{(1,1)} & \ldots &  {K}^{(1,r_1)}\\
\vdots & \null & \vdots\quad \\  {K}^{(r_1,1)} & \ldots &
 {K}^{(r_1,r_1)}\end{array}\right),
\label{KKbar}\end{equation}
with the matrices $ \overline{K}^{(\nu,\mu)}$ and $ {K}^{(\nu,\mu)}$ being given as
\begin{equation}
\overline{K}^{(\nu,\mu)}=
\sum_{j=0}^{s_\nu-1}\sum_{l=0}^{s_\mu-1}\frac{(-1)^l(j+l)!}{j!l!}
\frac{H^{(\nu)}_{-j}\overline{Q}_l^{(\nu,\mu)}}{(\kappa_\mu-\overline{\kappa}_\nu)^{j+l+1}},\quad
K^{(\nu,\mu)}= \sum_{l=0}^{s_\nu-1}\sum_{j=0}^{s_\mu-1}\frac{(-1)^l(l+j)!}{l!j!}
\frac{{Q}_l^{(\nu,\mu)}H^{(\mu)}_{j}}{(\overline{\kappa}_\nu-\kappa_\mu)^{l+j+1}}.
\label{Kblocks}\end{equation}
Here $\{H^{(\nu)}_{-s_\nu+1},\ldots,H^{(\nu)}_{s_\nu-1}\}$ is the basis  for the space
of $s_\nu\times s_\nu$-dimensional Toeplitz matrices, i.e., $(H^{(\nu)}_j)_{\alpha,\beta}
\equiv \delta_{\alpha,\beta-j}$.  The nonzero elements of matrices
$\overline{Q}_l^{(\nu,\mu)}$ and $Q_l^{(\nu,\mu)}$ are defined as the inner
products between the $p$-vectors from the blocks with indices $\nu$ and $\mu$:
\begin{equation}
\overline{Q}_l^{(\nu,\mu)} \equiv \left(\begin{array}{c} \langle\overline{p}^{(\nu)}_1|\\
\vdots\\ \langle\overline{p}^{(\nu)}_{s_\nu}|
\end{array}\right)(0,\ldots,0,|p^{(\mu)}_1\rangle,\ldots,|p^{(\mu)}_{s_\mu-l}\rangle),\quad
Q_l^{(\nu,\mu)} \equiv \left(\begin{array}{c} 0\\ \vdots\\0 \\ \langle\overline{p}^{(\nu)}_1|\\
\vdots\\ \langle\overline{p}^{(\nu)}_{s_\nu-l}|
\end{array}\right)(|p^{(\mu)}_1\rangle,\ldots,|p^{(\mu)}_{s_\mu}\rangle).
\label{QQbar}\end{equation}
\end{lem}

\noindent{\bf Remark 1 } In the case of a single pair of zeros $(k_1, \overline{k}_1)$,
we simply replace $\kappa_\mu \:(\overline{\kappa}_\mu)$ and $\kappa_\nu
\:(\overline{\kappa}_\nu)$ in formula (\ref{KKbar}) by $k_1 \: (\overline{k}_1)$.

\noindent{\bf Remark 2 } In the case of the involution (\ref{invol}) property, we have the obvious
relations:
\[
\overline{\kappa}_\nu = \kappa_\nu^*,\quad\langle \overline{p}^{(\nu)}_j| =
|p^{(\nu)}_j\rangle^\dagger,\quad \overline{D}_\nu(k) =
D^\dagger_\nu(k^*),\quad\overline{K}^{(\nu,\mu)} =
\left(K^{(\mu,\nu)}\right)^\dagger.
\]
\medskip

\noindent{\bf Proof} We only need to prove that the $|\overline{q}\rangle$ and
$\langle q|$ vectors in soliton matrices (\ref{S82})-(\ref{BLK2}) are related to
the $|p\rangle$ and $\langle \overline{p}|$ vectors by
\begin{equation}
(|\overline{q}^{(1)}_{s_1}\rangle,\ldots,|\overline{q}^{(1)}_{1}\rangle,\ldots,
|\overline{q}^{(r_1)}_{s_{r_1}}\rangle,\ldots,|\overline{q}^{(r_1)}_{1}\rangle)\overline{\cal
K} =- (|p^{(1)}_{1}\rangle,\ldots,|p^{(1)}_{s_1}\rangle,\ldots,
|p^{(r_1)}_{1}\rangle,\ldots,|p^{(r_1)}_{s_{r_1}}\rangle),
\label{2qBLK}\end{equation}
and
\begin{equation}
{\cal K} \left(\begin{array}{c} \langle q_{s_1}^{(1)}| \\  \vdots \\  \langle q_1^{(1)}| \\  \vdots
\\ \langle q_{s_{r_1}}^{(r_1)}| \\ \vdots \\ \langle q_1^{(r_1)}|\end{array}\right)
=-\left(\begin{array}{c} \langle \overline{p}_{1}^{(1)}| \\  \vdots \\  \langle \overline{p}_{s_1}^{(1)}| \\ \vdots
\\ \langle \overline{p}_1^{(r_1)}| \\ \vdots \\ \overline{p}_{s_{r_1}}^{(r_1)}|\end{array}\right),
\label{2qBLK2}\end{equation}
where matrices ${\cal K}$ and $\overline{\cal K}$ are as given in Eq. (\ref{KKbar}).
We will give the proof only for Eq. (\ref{2qBLK}), as the proof for
(\ref{2qBLK2}) is similar. Note that in the case of involution (\ref{invol}),
Eq. (\ref{2qBLK2}) is equivalent to (\ref{2qBLK}) by taking the Hermitian.

To prove Eq. (\ref{2qBLK2}), we consider the corresponding expression
(\ref{S82})-(\ref{BLK2}) for $\Gamma(k)$:
\begin{equation}
\Gamma(k) = I +
\sum_{\nu=1}^{r_1}(|\overline{q}^{(\nu)}_{s_\nu}\rangle,\ldots,|\overline{q}^{(\nu)}_{1}\rangle)
\overline{D}_{\nu}(k)\left(\begin{array}{c} \langle\overline{p}^{(\nu)}_1|\\ \vdots\\
\langle\overline{p}^{(\nu)}_{s_\nu}| \end{array}\right).
\label{2BLK}\end{equation}
We need to determine the $|\overline{q}\rangle$-vectors from Eq. (\ref{S14}).
Note that the $l$-th row in the $\mu$-system (\ref{S14}) can be written as
\begin{equation}
\left[\Gamma(\kappa_\mu),\frac{1}{1!}\frac{\mathrm{d}\Gamma}{\mathrm{d}k}(\kappa_\mu),\ldots,
\frac{1}{(l-1)!}\frac{\mathrm{d}^{l-1}\Gamma}{\mathrm{d}k{}^{l-1}}(\kappa_\mu)\right]
\left(\begin{array}{c}|p^{(\mu)}_{l}\rangle\\
\vdots\\|p^{(\mu)}_1\rangle\end{array}\right) = 0
\end{equation}
for each $1\le \mu\le r_1$.
When the $\Gamma(k)$ expression (\ref{2BLK}) is substituted into the above equation,
we get
\[
\sum_{\nu=1}^{r_1}
(|\overline{q}^{(\nu)}_{s_\nu}\rangle,\ldots,|\overline{q}^{(\nu)}_{1}\rangle)
\left\{ \overline{D}_\nu(\kappa_\mu)\left(\begin{array}{c} \langle\overline{p}^{(\nu)}_1|\\ \vdots\\
\langle\overline{p}^{(\nu)}_{s_\nu}| \end{array}\right)|p^{(\mu)}_l\rangle +
\frac{1}{1!}\frac{\mathrm{d}\overline{D}_\nu}{\mathrm{d} k}(\kappa_\mu)\left(\begin{array}{c} \langle\overline{p}^{(\nu)}_1|\\ \vdots\\
\langle\overline{p}^{(\nu)}_{s_\nu}|
\end{array}\right)|p^{(\mu)}_{l-1}\rangle\right.
\]
\begin{equation}
\left. +\ldots +
\frac{1}{(l-1)!}\frac{\mathrm{d}^{l-1}\overline{D}_\nu}{\mathrm{d}
k^{l-1}}(\kappa_\mu)\left(\begin{array}{c}
 \langle\overline{p}^{(\nu)}_1|\\ \vdots\\ \langle\overline{p}^{(\nu)}_{s_\nu}|
\end{array}\right)|p^{(\mu)}_{1}\rangle\right\} = - |p^{(\mu)}_l\rangle.
\label{lthrow}\end{equation}
The derivatives of $\overline{D}_\nu(\kappa_\mu)$ can be easily computed:
\begin{equation}
\frac{\mathrm{d}^l\overline{D}_\nu}{\mathrm{d}k^l}(\kappa_\mu) =
\sum_{j=0}^{s_\nu-1}\frac{(-1)^l(j+l)!}{j!}\frac{H^{(\nu)}_{-j}}{(\kappa_\mu-\overline{\kappa}_\nu)^{j+l+1}}.
\label{sub2}\end{equation}
Now it is straightforward to verify that all equations of the type (\ref{lthrow})
can be united in a
single matrix equation (\ref{2qBLK})
by padding some columns in the summations of (\ref{lthrow})
by zeros, precisely as it is done in the definition (\ref{QQbar})
of $\overline{Q}^{(\nu,\mu)}$.
As a result we arrive at the relation (\ref{2qBLK}) between
$|\overline{q}\rangle$ and $|p\rangle$ vectors, where the matrix $\overline{\cal K}$ is precisely
as defined in Lemma \ref{lem4}. Q.E.D.

\subsection{Two special cases}
The soliton matrices derived above
reproduce all previous results as special cases.
Previous results were obtained in two special cases:
several pairs of Riemann-Hilbert zeros with equal geometric and algebraic
multiplicities \cite{Kawata}, and a single pair of elementary Riemann-Hilbert zeros \cite{SIAM}.
In the first case, suppose the geometric and algebraic multiplicities of
$n$ pairs of Riemann-Hilbert zeros
$\{(k_j,\overline{k}_j), 1\le j\le n\}$ are $\{r^{(j)}, 1\le j\le n\}$ respectively.
Then the soliton matrices have been given before \cite{Kawata} (see also appendix B in
Ref.~\cite{TMF1}) as:
\begin{equation}
\Gamma = I - \sum_{i,j=1}^n\sum_{m=1}^{r^{(i)}}\sum_{l=1}^{r^{(j)}} \frac{| v^{(m)}_i
\rangle\left(F^{-1}\right)_{im,jl}\langle \overline{ v}{}^{(l)}_j |}{k -
\overline{k}_j} ,\quad \Gamma^{-1} = I +
\sum_{i,j=1}^n\sum_{m=1}^{r^{(i)}}\sum_{l=1}^{r^{(j)}} \frac{ | v^{(l)}_j
\rangle\left(F^{-1}\right)_{jl,im} \langle \overline{ v}{}^{(m)}_i |}{k - {k}_j},
\label{exmpl1}
\end{equation}
where $r^{(j)}$ vectors $\{|v_j^{(l)}\rangle, 1\le l\le r^{(j)}\}$ and
$\{\langle\overline{v}_j^{(l)}|, 1\le l\le r^{(j)}\}$ are in the kernels
of $\Gamma(k_j)$ and $\Gamma^{-1}(\overline{k}_j)$ respectively:
\begin{equation}
\Gamma(k_j)| v^{(l)}_j \rangle = 0,\quad  \langle \overline{ v}{}^{(l)}_j |
\Gamma^{-1}(\overline{k}_j) = 0,\quad l=1,...,r^{(j)},
\label{exmpl3}\end{equation}
and
\begin{equation}
F_{im,jl} = \frac{\langle \overline{ v}{}^{(m)}_i | v^{(l)}_j \rangle }{k_j -
\overline{k}_i}.
\label{exmpl2}\end{equation}
Moreover,
\[
\det \Gamma = \prod_{j=1}^n\left(\frac{k-k_j}{k-\overline{k}_j}\right)^{r^{(j)}}.
\]
The above special soliton matrices
can be easily retrieved from the general soliton matrices
(\ref{mainrepr})-(\ref{QQbar}) of lemma~\ref{lem4}. Indeed,
in this special case, the block sequence of a pair of zeros $(k_j, \overline{k}_j)$
is $r^{(j)}$ consecutive 1's.  Thus $s_\nu=1$ for all $\nu$'s. Consequently,
matrices $D_\nu$ and $\overline{D}_\nu$ in Eq. (\ref{DDbar}) have dimension 1. In addition,
matrices $K^{(\nu,\mu)}$ and $\overline{K}^{(\nu,\mu)}$ in Eq. (\ref{Kblocks})
also have dimension 1, and the summations in their definitions can be dropped since $l=0$
and $j=0$ there. Hence, we get
\[\overline{K}^{(\nu,\mu)} =\left(K^{(\mu,\nu)}\right)^\dagger=
\frac{\langle\overline{p}^{(\nu)}_1|p^{(\mu)}_1\rangle}{\kappa_\mu-\overline{\kappa}_\nu},\]
see (\ref{QQbar}). Relating $|p\rangle$-vectors
$\{|p_1^{(\nu)}\rangle, 1+\sum_{l=1}^{j-1}r^{(l)}
\le \nu\le \sum_{l=1}^{j}r^{(l)}\}$
to $\{|v_j^{(l)}\rangle, 1\le l \le r^{(j)}\}$
and $\{\langle p_1^{(\nu)}|, 1+\sum_{l=1}^{j-1}r^{(l)}
\le \nu\le \sum_{l=1}^{j}r^{(l)}\}$
to $\{\langle v_j^{(l)}|, 1\le l \le r^{(j)}\}$ for each $j=1, \dots, n$, and recalling
the definition (\ref{kappa}) of $\kappa$'s, we readily find that
our general representation (\ref{mainrepr}) reduces to (\ref{exmpl1}).
We note by passing that the soliton matrices
(\ref{exmpl1})-(\ref{exmpl2}) cover the case of simple zeros, where there is just
one vector in each kernel in (\ref{exmpl3}).

Our second example is a single pair of elementary higher-order zeros. A
higher-order zero is called elementary if its geometric multiplicity is 1 \cite{SIAM}.
This case has been extensively studied in the literature before (see
Refs.~\cite{Wadati,OptLett,Nathalie,SIAM}) for different integrable PDEs. The
soliton matrices having similar representation as (\ref{mainrepr})-(\ref{QQbar})
for this case were derived in our previous publication \cite{SIAM}. The only
difference between that paper's representation and the present one
(\ref{mainrepr})-(\ref{QQbar}) is the definition of the matrices ${\cal K}$ and
$\overline{\cal K}$. However, in this special case, these matrices have just one
block each, i.e., $K^{(1,1)}$ and $\overline{K}^{(1,1)}$, since there
is just one $\nu$-block in the soliton matrices. By comparison of both definitions
one can easily establish their equivalence.

\subsection{Invariance properties of soliton matrices}
\label{invarian}
In this subsection, we discuss the invariance properties of soliton matrices.
When the soliton matrix is in the product representation (\ref{S6}) for a single pair of zeros,
the invariance property means that one can choose any $r_1$ linearly independent
vectors in the kernels of $\Gamma(k_1)$ and $\Gamma^{-1}(\overline{k}_1)$, or more generally,
one can choose any $r_l \; (1\le l \le n)$ linearly independent
vectors in the kernels of $(\Gamma\chi_1^{-1}\dots \chi_{l-1}^{-1})(k_1)$
and $(\chi_{l-1}\dots \chi_1\Gamma^{-1})(\overline{k}_1)$, and
the soliton matrix remains invariant.
In other words, given the soliton matrix $\Gamma(k)$ for
a fixed set of $r_l$ linearly independent
vectors $|v_{il}\rangle \;(1\le i \le r_l)$
in the kernels of $(\Gamma\chi_1^{-1}\dots \chi_{l-1}^{-1})(k_1)$
and another fixed set of $r_l$ linearly independent
vectors $\langle \overline{v}_{il}| \;(1\le i \le r_l)$
in the kernels of $(\chi_{l-1}\dots \chi_1\Gamma^{-1})(\overline{k}_1)$,
new sets of vectors
\begin{equation}\label{trans1}
\left[|\widetilde{v}_{1l}\rangle, |\widetilde{v}_{2l}\rangle, \dots, |\widetilde{v}_{r_l,l}\rangle\right]
=\left[|{v}_{1l}\rangle, |{v}_{2l}\rangle, \dots, |{v}_{r_l,l}\rangle\right]B,
\end{equation}
and
\begin{equation}\label{trans2}
\left[\begin{array}{c} \langle \widetilde{\overline{v}}_{1l}| \\
            \langle \widetilde{\overline{v}}_{2l}| \\
\vdots \\  \langle \widetilde{\overline{v}}_{r_l,l}| \end{array}\right]
=\overline{B}\left[\begin{array}{c} \langle \overline{v}_{1l}| \\
            \langle \overline{v}_{2l}| \\
\vdots \\  \langle \overline{v}_{r_l,l}| \end{array}\right],
\end{equation}
where $B$ and $\overline{B}$ are arbitrary $k$-independent non-degenerate
$r_l\times r_l$ matrices, give the same soliton matrix $\Gamma(k)$. This invariance
property is obvious from definitions (\ref{S5}) for projector matrices. Note that
the invariance transformations (\ref{trans1})-(\ref{trans2}) are the most general
automorphisms of the respective kernels (null spaces) $(\Gamma\chi_1^{-1}\dots
\chi_{l-1}^{-1})(k_1)$ and $(\chi_{l-1}\dots \chi_1\Gamma^{-1})(\overline{k}_1)$.

Now let us determine the total number ${\cal N}_\mathrm{free}$ of free complex
parameters characterizing the higher-order soliton solution. For a single pair of
the  higher-order zeros $(k_1,\overline{k}_1)$ in the case with no involution, it
is given by the total number ${\cal N}_\mathrm{tot}$ $(=2N\sum_{l=1}^n r_l+2)$ of
all complex constants in all the linearly independent vectors in the above null
spaces and the pair of zeros $(k_1, \overline{k}_1)$, minus the total number ${\cal
N}_\mathrm{inv}$ $(=2\sum_{l=1}^n r_l^2)$ of the free parameters in the invariance
matrices (\ref{trans1})-(\ref{trans2}). Thus, in the case with no involution, we
have
\begin{equation}
{\cal N}_\mathrm{free} \equiv {\cal N}_\mathrm{tot} - {\cal N}_\mathrm{inv} =
2N\sum_{l=1}^n r_l + 2  - 2\sum_{l=1}^n r_l^2.
\label{Nfree}\end{equation}
Note that the total number of $|v\rangle$ or $\langle \overline{v}|$ vectors in the
product representation (\ref{S6}), given by the
sum $\sum_{l=1}^n r_l$, is equal to the algebraic order of the pair of zeros ($k_1,
\overline{k}_1)$. In the case of the involution (\ref{invol}),
the number ${\cal N}_\mathrm{free}$ is reduced by half.
When the soliton matrices have several pairs of
zeros as in the product representation (\ref{gengam}), the invariance property is
similar, and the total number of free soliton parameters is given by the sum of the r.h.s of
formula (\ref{Nfree}) for all distinct pairs of zeros.

By analogy, the invariance properties for the summation representation
(\ref{mainrepr}) of the soliton matrices are defined as preserving the form of the
soliton matrices as well as the equations defining the $|p\rangle$- and $\langle
\overline{p}|$-vectors (\ref{S14})-(\ref{S15}). The equations defining the
transformations between different sets of $p$-vectors of the same invariance class
must be linear, since all the sets of $p$-vectors in the invariance class satisfy
equations (\ref{S14})-(\ref{S15}) for a {\it fixed} soliton matrix -- i.e. the
invariance transformations are a subset of transformations between solutions to a
set of {\it linear} equations. Thus the most general form of the invariance is
given by two linear transformations
--- one for $|p\rangle$-vectors and one for $\langle \overline{p}|$-vectors:
\begin{equation}\label{transnew1}
(|\widetilde{p}^{(1)}_1\rangle,\ldots,|\widetilde{p}^{(1)}_{s_1}\rangle,
\ldots,|\widetilde{p}^{(r_1)}_1\rangle,\ldots,|\widetilde{p}^{(r_1)}_{s_{r_1}}\rangle)
=(|p^{(1)}_1\rangle,\ldots,|p^{(1)}_{s_1}\rangle,\ldots,|p^{(r_1)}_1\rangle,\ldots,|p^{(r_1)}_{s_{r_1}}\rangle)B,
\end{equation}
and
\begin{equation}\label{transnew2}
\left(\begin{array}{c}
\langle\widetilde{\overline{p}}^{(1)}_1|\\ \vdots\\ \langle\widetilde{\overline{p}}^{(1)}_{s_1}| \\
\vdots \\ \langle\widetilde{\overline{p}}^{(r_1)}_1| \\ \vdots \\
\langle\widetilde{\overline{p}}^{(r_1)}_{s_{r_1}}|\end{array}\right)
=\overline{B}\left(\begin{array}{c}
\langle\overline{p}^{(1)}_1|\\ \vdots\\ \langle\overline{p}^{(1)}_{s_1}| \\
\vdots \\ \langle\overline{p}^{(r_1)}_1| \\ \vdots \\
\langle\overline{p}^{(r_1)}_{s_{r_1}}|\end{array}\right).
\end{equation}
Different from the product representation of the soliton matrices, the
transformation matrices $B$ and $\overline{B}$ in Eqs. (\ref{transnew1}) and
(\ref{transnew2}) can not be arbitrary in order to keep the soliton matrices
(\ref{mainrepr}) and equations (\ref{S14})-(\ref{S15}) invariant. Let us call
such matrices $B$ and $\overline{B}$ which keep the soliton matrices
(\ref{mainrepr}) invariant as invariance matrices. The forms of invariance matrices
can be determined most easily by considering the invariance of equations
(\ref{S14})-(\ref{S15}).

Recall from Lemma \ref{lem3} that all $|p\rangle$ vectors in the soliton matrix
(\ref{mainrepr}) satisfy the equation
\begin{equation}\label{invar}
{\bf \Gamma_{\bf B}} \left(\begin{array}{c} |p_1^{(1)}\rangle \\ \vdots \\ |p_{s_1}^{(1)}\rangle \\ \vdots \\
|p_1^{(r_1)}\rangle \\ \vdots  \\ |p_{s_{r_1}}^{(r_1)}\rangle \end{array}\right)=0,
\quad \quad\quad
{\bf \Gamma_{\bf B}} \equiv \left(\begin{array}{ccc}{\bf \Gamma}_1(\kappa_1) & \null & 0\\
\null & \ddots & \null \\ 0 & \null & {\bf \Gamma}_{r_1}(\kappa_{r_1}) \end{array}\right),
\end{equation}
where ${\bf \Gamma}_\nu(\kappa_\nu)$ is the lower-triangular Toeplitz matrix defined in
Eq. (\ref{S14}). The matrix $B$ is an invariance matrix if and only if
the above equation is still satisfied when the $|p\rangle$ vectors in Eq. (\ref{invar})
are replaced by the transformed vectors $|\widetilde{p}\rangle$ in Eq. (\ref{transnew1}),
and the resulting matrices ${\cal K}$ and $\overline{\cal K}$ are non-degenerate
[see Eq. (\ref{mainrepr})].
Note that the transformation (\ref{transnew1}) can be rewritten in the following form:
\begin{equation}\label{ptilde}
\left(\begin{array}{c} |\widetilde{p}_1^{(1)}\rangle \\ \vdots \\ |\widetilde{p}_{s_1}^{(1)}\rangle \\ \vdots \\
|\widetilde{p}_1^{(r_1)}\rangle \\ \vdots  \\ |\widetilde{p}_{s_{r_1}}^{(r_1)}\rangle \end{array}\right)
=B^T \left(\begin{array}{c} |p_1^{(1)}\rangle \\ \vdots \\ |p_{s_1}^{(1)}\rangle \\ \vdots \\
|p_1^{(r_1)}\rangle \\ \vdots  \\ |p_{s_{r_1}}^{(r_1)}\rangle \end{array}\right),
\end{equation}
where the superscript "$T$" represents the transpose of a matrix.
Since the original $|p\rangle$ vectors can be chosen arbitrarily
(the matrix $\bf \Gamma_{\bf B}$ is determined subsequently from
these $|p\rangle$ vectors as well as the $\langle \overline{p}|$ vectors), in order for
the above $|\widetilde{p}\rangle$ vectors (\ref{ptilde}) to satisfy Eq. (\ref{invar}) as well,
the necessary and sufficient condition is that
${\bf \Gamma}_{\bf B}$ and $B^T$ commute, i.e.,
\begin{equation} \label{comm1}
{\bf \Gamma}_{\bf B} \cdot B^T =B^T \cdot {\bf \Gamma}_{\bf B},
\end{equation}
and $B$ is non-degenerate.
The requirement for the non-degeneracy of $B$ is needed in order for
the resulting matrices $\widetilde{\cal K}$ and $\widetilde{\overline{\cal K}}$
to be non-degenerate [see Eq. (\ref{Ktrans})].
Similarly, we can show that the matrix $\overline{B}$ in
Eq. (\ref{transnew2}) is an invariance matrix if and only if
$\overline{\bf \Gamma}_{\bf B}$ and $\overline{B}^T$ commute,
\begin{equation} \label{comm2}
\overline{\bf \Gamma}_{\bf B}\cdot \overline{B}^T = \overline{B}^T \cdot \overline{\bf \Gamma}_{\bf B},
\end{equation}
and $\overline{B}$ is non-degenerate. Here the block-diagonal matrix $\overline{\bf \Gamma}_{\bf B}$ is
\begin{equation} \label{GammaBbar}
\overline{\bf \Gamma}_{\bf B} \equiv \left(\begin{array}{ccc}\overline{\bf \Gamma}_1(\kappa_1) & \null & 0\\
\null & \ddots & \null \\ 0 & \null & \overline{\bf \Gamma}_{r_1}(\kappa_{r_1}) \end{array}\right),
\end{equation}
and upper-triangular Toeplitz matrices $\overline{\bf \Gamma}_\nu(\kappa_\nu)$ have been defined in
Eq. (\ref{S15}).
Note that matrices ${\bf \Gamma}_{\bf B}$ and $\overline{\bf \Gamma}_{\bf B}$ have exactly
the same forms as $\overline{\cal D}(k)$ and ${\cal D}(k)$ respectively. Thus
invariance matrices $B^T$ and $\overline{B}^T$ commute with $\overline{\cal D}(k)$ and ${\cal D}(k)$
as well:
\begin{equation}\label{comm3old}
\overline{\cal D}(k)\cdot B^T=B^T\cdot \overline{\cal D}(k), \quad \quad
{\cal D}(k)\cdot \overline{B}^T=\overline{B}^T\cdot {\cal D}(k).
\end{equation}
In addition, since ${\cal D}^T$ has the same form as $\overline{\cal D}$, invariance matrices $B$
and $\overline{B}$ also commute with ${\cal D}$ and $\overline{\cal D}$:
\begin{equation}\label{comm3}
B\cdot {\cal D}(k) ={\cal D}(k)\cdot B, \quad \quad \overline{B}\cdot \overline{\cal D}(k) =
\overline{\cal D}(k)\cdot \overline{B}.
\end{equation}

The forms of these invariance matrices are easy to determine.
First of all, the commutability relations
(\ref{comm3}) demand that the invariance matrix $B$ has a block-diagonal form with
each block corresponding to a pair of zeros:
\begin{equation} \label{Bform}
B=\left(\begin{array}{cccc} B_1 & & & \\ & B_2 & & \\& & \ddots & \\ & & & B_{N_Z} \end{array} \right).
\end{equation}
Here $B_n$ is a square matrix associated with the $n$-th pair of zeros $(k_n, \overline{k}_n)$.
The form of each matrix $B_n$ is readily found to be
\begin{equation} \label{Bn}
B_n = \left(\begin{array}{ccc}  B_n^{(1,1)} & \ldots &  B_n^{(1,r_1^{(n)})}\\
\vdots & \null & \vdots\quad \\  B_n^{(r_1^{(n)},1)} & \ldots &
 B_n^{(r_1^{(n)},r_1^{(n)})}\end{array}\right),
\end{equation}
where $B_n^{(\nu,\mu)}$ is a $s_\nu^{(n)}\times s_\mu^{(n)}$ matrix of the following type:
\begin{subequations} \label{Bforms}
\begin{equation}
B_n^{(\nu,\mu)} =
\left(\begin{array}{cccccccc}
0 & \dots & 0 & b_1 & b_2 & \dots & b_{s_\nu^{(n)}-1} & b_{s_\nu^{(n)}} \\
0& \ddots & \ddots & 0 & b_1 & b_2 & \ddots & b_{s_\nu^{(n)}-1} \\
\vdots & \ddots & \ddots & \ddots & \ddots  & \ddots & \ddots & \vdots \\
0 & \ddots & \ddots & \ddots & \ddots & 0 &b_1 & b_2 \\
0 & \dots & \dots & \dots & \dots & \dots &  0 & b_1 \end{array} \right), \quad
\quad \nu \ge \mu,
\end{equation}
\begin{equation}
B_n^{(\nu,\mu)} =
\left(\begin{array}{ccccc}
c_1 & c_2 & \dots & c_{s_\mu^{(n)}-1} & c_{s_\mu^{(n)}} \\
0& c_1 & c_2 & \ddots & c_{s_\mu^{(n)}-1} \\
\vdots & 0 & \ddots & \ddots & \vdots \\
\vdots & \ddots & \ddots & \ddots & c_2 \\
\vdots & \ddots & \ddots & 0 &c_1 \\
\vdots & \ddots & \ddots & \ddots & 0 \\
\vdots & \ddots & \ddots & \ddots & \vdots \\
0 & \dots & \dots & \dots & 0
\end{array} \right), \quad \quad \nu \le \mu,
\end{equation}
\end{subequations}
$s_1^{(n)} \ge s_2^{(n)} \ge \dots \ge s_{r_1^{(n)}}^{(n)}$ is the block sequence
of zeros $(k_n, \overline{k}_n)$ as in Lemma \ref{lem2} (see Definition
\ref{def2}), and $b_j, c_j$ are arbitrary complex constants which are generally different in
different submatrices $B_n^{(\nu,\mu)}$. The invariance matrix $\overline{B}$ has the
form of $B^T$.

The above forms (\ref{Bn})-(\ref{Bforms})
of the invariance matrices $B_n$  and $\overline{B}_n$ follow
immediately from the following argument. Consider, for instance, the matrix $B_n$.
The commutability relation with the part of the matrix ${\cal D}(k)$ corresponding
to the $n$-th pair of zeros, i.e.,
${\cal D}^{(n)}(k) =\mathrm{diag}[D^{(n)}_1(k),...,D^{(n)}_{r^{(n)}_1}(k)]$
where matrices $D^{(n)}_\nu(k)$ are given by Eq. (\ref{Ds}),
produces the following set of independent matrix equations
\begin{equation}
D^{(n)}_\nu(k) B^{(\nu,\mu)}_n = B^{(\nu,\mu)}_n D^{(n)}_\mu(k), \quad \nu,\mu =
1,...,r_1^{(n)}.
\label{set1}\end{equation}
For $\nu=\mu$, the above equations are equivalent to the commutability conditions
for the single elementary higher-order zero considered in Ref.~\cite{SIAM}, thus
the form (\ref{Bforms})
for the diagonal blocks $B^{(\nu,\nu)}_n$ follows accordingly.
Consider now the case when $\nu>\mu$ (the other case can be considered
similarly). We have then $s_\nu^{(n)} \le s^{(n)}_\mu$, and the square matrix
$D^{(n)}_\mu(k)$ contains the matrix $D^{(n)}_\nu(k)$ in its lower right corner
[consult the definition (\ref{Ds})]. It is easy to conclude, first of all, that the first
$\mu-\nu$ columns of the matrix $B^{(\nu,\mu)}_n$ are identically zero, otherwise on
the r.h.s. of equation (\ref{set1}) we would have higher powers of $(k-k_n)^{-1}$
than the highest power of this quantity on the l.h.s.. Then if we denote the
non-zero part of $B^{(\nu,\mu)}_n$ as $\hat{B}^{(\nu)}_n$, the condition
(\ref{set1}) becomes
\[
D^{(n)}_\nu(k)\hat{B}^{(\nu)}_n = \hat{B}^{(\nu)}_n D^{(n)}_\nu(k),
\]
which is equivalent to the one considered above in the case of $\mu = \nu$. Thus
the form (\ref{Bforms}a)
for the  off-diagonal blocks of the invariance matrix $B^{(\nu,\mu)}_n$
follows as well. Q.E.D.
\bigskip

From the above explicit expressions (\ref{Bform})-(\ref{Bforms})
for invariance matrices in the summation representation (\ref{mainrepr}),
it is easy to see that the
total number ${\cal N}_\mathrm{inv}$ of free complex constants in these
invariance matrices
coincides with that in the product representation (\ref{S6}) and (\ref{gengam})
[see Eq. (\ref{Nfree})].
Indeed consider  for simplicity just a single
pair of zeros. In the case with no involution (\ref{invol}),
the total number ${\cal N}_\mathrm{inv}$ of free complex
constants in the invariance matrices (\ref{Bform})-(\ref{Bforms}) is
\[
{\cal N}_\mathrm{inv} = 2\sum_{\nu=1}^{r_1} \left( 2r_1-2\nu +1\right)
s_{r_1-\nu+1} = 2\sum_{\mu=1}^{r_1}(2\mu-1)s_\mu
\]
\begin{equation} \label{number}
= 2\left( n\sum_{\mu=1}^{r_n}(2\mu-1) + (n-1)\sum_{\mu=r_n+1}^{r_{n-1}}(2\mu-1) +
... +\sum_{\mu=r_2+1}^{r_1}(2\mu-1) \right) = 2\sum_{l=1}^n r_l^2,
\end{equation}
which is exactly the same as that in Eq. (\ref{Nfree}) for ${\cal N}_\mathrm{inv}$.
Here we have used the fact that the numbers of blocks with sizes $[1,2,3,...,n]$
are given by the differences of the ranks $[r_1-r_2,r_2-r_3,...,r_{n-1}-r_n,r_n]$
(see end of Sec. \ref{3B}). This result is not surprising since the invariance
properties of the soliton matrices in the summation representation originate from
the invariance properties in the product representation, that is why the respective
invariance matrices have the same total number of free parameters. Consequently,
the total number of free complex parameters in the summation representation
(\ref{mainrepr}) is the same as in the product representation, in the case with no
involution for a single pair of zeros  it is given by the same Eq. (\ref{Nfree}).

Invariance matrices have many important properties. These include (i) the identity
matrix $I$ is an invariance matrix; (ii) if $B$ is an invariance matrix, so is
$cB$, where $c$ is any non-zero complex constant; (iii) if $B$ is an invariance
matrix, so is $B^{-1}$; (iv) if $B_1$ and $B_2$ are two invariance matrices, so are
$B_1\pm B_2$ and $B_1\cdot B_2$. In the former case, $B_1\pm B_2$ should be
non-degenerate.

Lastly, we note that if matrices $B$ and $\overline{B}$ satisfy the
commutability relations (\ref{comm3}), the transformations (\ref{transnew1})
and (\ref{transnew2}) indeed keep the soliton matrices (\ref{mainrepr}) invariant.
The proof uses the fact that under the transformation (\ref{transnew1})
where $B$ is an invariance matrix (the $\langle \overline{p}|$ vectors are held fixed),
matrices ${\cal K}$ and $\overline{\cal K}$ are transformed to
\begin{equation} \label{Ktrans}
\widetilde{\cal K}={\cal K}B, \quad \quad
\widetilde{\overline{\cal K}}=\overline{\cal K} B
\end{equation}
respectively.
Similarly, under the transformation (\ref{transnew2}) while keeping the $|p\rangle$ vectors
fixed, matrices ${\cal K}$ and $\overline{\cal K}$ are transformed to
\begin{equation} \label{Ktrans2}
\widetilde{\cal K}=\overline{B} {\cal K}, \quad \quad \widetilde{\overline{\cal
K}}=\overline{B}\,\overline{\cal K}.
\end{equation}
For a single pair of elementary zeros, these facts
have been proved in \cite{SIAM}. The proof for the present general case is given below.
Since the proofs for Eqs. (\ref{Ktrans}) and (\ref{Ktrans2}) are similar, we only consider
Eq. (\ref{Ktrans}).

To prove the transformation (\ref{Ktrans}), we need to recall how matrices
${\cal K}$ and $\overline{\cal K}$ are obtained. The matrix $\overline{\cal K}$
is derived from Eq. (\ref{invar}). Comparing this equation with (\ref{2qBLK}),
we find that
\[ ({\bf \Gamma}_{\bf B}-I)\left(\begin{array}{c} |p_1^{(1)}\rangle \\ \vdots \\ |p_{s_1}^{(1)}\rangle \\ \vdots \\
|p_1^{(r_1)}\rangle \\ \vdots  \\ |p_{s_{r_1}}^{(r_1)}\rangle \end{array}\right)
=\overline{\cal K}^T
\left(\begin{array}{c} |\overline{q}_{s_1}^{(1)}\rangle \\ \vdots \\ |\overline{q}_{1}^{(1)}\rangle \\ \vdots \\
|\overline{q}_{s_{r_1}}^{(r_1)}\rangle \\ \vdots  \\ |\overline{q}_{1}^{(r_1)}\rangle \end{array}\right).
\]
Now under the transformation (\ref{transnew1}), i.e., (\ref{ptilde}), and
recalling that $B^T$ and ${\bf \Gamma}_{\bf B}-I$ commute, we readily find that
$\widetilde{\overline{\cal K}}^T=B^T \overline{\cal K}^T$, thus
$\widetilde{\overline{\cal K}}=\overline{\cal K} B$.
As about the matrix ${\cal K}$, it is derived from the equation
\[\left( \langle \overline{p}_1^{(1)}|, \dots, \langle \overline{p}_{s_1}^{(1)}|,
 \dots, \langle \overline{p}_1^{(r_1)}|, \dots, \langle \overline{p}_{s_{r_1}}^{(r_1)}|\right)
\overline{\bf \Gamma}_{\bf B}=0, \]
where $\overline{\bf \Gamma}_{\bf B}$ is given by Eqs. (\ref{S14}) and (\ref{GammaBbar}).
Recall that $\Gamma^{-1}(k)$ is given by Eq. (\ref{S82}), i.e.,
\[\Gamma^{-1}(k)=
I+(|p^{(1)}_1\rangle,\ldots,|p^{(1)}_{s_1}\rangle,\ldots,|p^{(r_1)}_1\rangle,\ldots,|p^{(r_1)}_{s_{r_1}}\rangle)
{\cal D}(k)
\left(\begin{array}{c} \langle q_{s_1}^{(1)}| \\  \vdots \\  \langle q_1^{(1)}| \\  \vdots
\\ \langle q_{s_{r_1}}^{(r_1)}| \\ \vdots \\ \langle q_1^{(r_1)}|\end{array}\right).
\]
Under the transformation (\ref{transnew1}), noting that $B$ and ${\cal D}$ commute
[see Eq. (\ref{comm3})], we readily find that $\widetilde{\cal K}={\cal K}B$.
Thus (\ref{Ktrans}) holds.

Because of Eq. (\ref{Ktrans}) and the commutability relation (\ref{comm3}),
we see that soliton matrices $\Gamma(k)$ and $\Gamma^{-1}(k)$ in Eq. (\ref{mainrepr})
indeed remain invariant under the transformation (\ref{transnew1}).
Analogously, these soliton matrices are also invariant under the transformation
(\ref{transnew2}) if matrix $\overline{B}$ is an invariance matrix.
In the case of involution (\ref{invol}), transformations
(\ref{transnew1}) and (\ref{transnew2}) need to be performed simultaneously
since $|p\rangle$ and $\langle \overline{p}|$ vectors are related
by the Hermitian operation. Under these combined transformations,
matrix ${\cal K}$ transforms to $\widetilde{\cal K}=\overline{B}{\cal K}B$, thus
soliton matrices (\ref{mainrepr}) remain invariant as well.

The invariance matrices can be used to reduce the free parameters in soliton
solutions to a minimum as we have done above
[see Eq. (\ref{Nfree})]. It is also needed
to classify the general evolution of soliton matrices (see the next subsection).

\subsection{Spatial and temporal evolutions of soliton matrices}
Finally, we derive the $(x,t)$-dependence of the vector parameters which enter the
soliton matrix (\ref{mainrepr}). The idea is similar to that in
the derivation of equations (\ref{RH8}) in section \ref{secRH}.
Our starting point is the fact that the soliton matrix $\Gamma(k,x,t)$
satisfies equations (\ref{RH3})-(\ref{RH4}) with potentials $U(k,x,t)$ and
$V(k,x,t)$:
\begin{subequations}
\label{S27}\begin{eqnarray}
\partial_x \Gamma(k,x,t) &=& \Gamma(k,x,t)\Lambda(k) +
U(k,x,t)\Gamma(k,x,t),
\label{S27a}\\
\partial_t \Gamma(k,x,t) &=& \Gamma(k,x,t)\Omega(k) +
V(k,x,t)\Gamma(k,x,t).
\label{S27b}\end{eqnarray}
\end{subequations}
First we need to find the equations for the triangular block-Toeplitz matrices
${\bf\Gamma}_\nu$ and $\overline{\bf\Gamma}_\nu$. To this goal one needs to
differentiate equations (\ref{S27}) with respect to $k$ up to the $(s_\nu-1)$-th
order. It is easy to check that the equations for
${\bf\Gamma}_\nu$ have the same form as equations (\ref{S27}):
\begin{subequations}
\label{S28}
\begin{eqnarray}
\partial_x {\bf\Gamma}_\nu(k,x,t) &=& {\bf\Gamma}_\nu(k,x,t){\bf\Lambda}_\nu(k) +
{\bf U}_\nu(k,x,t){\bf\Gamma}_\nu(k,x,t),
\label{S28a}\\
\partial_t {\bf\Gamma}_\nu(k,x,t) &=& {\bf\Gamma}_\nu(k,x,t){\bf\Omega}_\nu(k) +
{\bf V}_\nu(k,x,t){\bf\Gamma}_\nu(k,x,t).
\label{S28b}\end{eqnarray}
\end{subequations}
Here ${\bf\Lambda}_\nu$, ${\bf\Omega}_\nu$, ${\bf U}_\nu$, and ${\bf V}_\nu$ are
lower-triangular block-Toeplitz  matrices:
\begin{equation}
{\bf\Lambda}_\nu \equiv \left(\begin{array}{cccc}
\Lambda&0 & \ldots& \quad 0\\
\frac{1}{1!}\frac{\text{d}}{\text{d}k}\Lambda
&\ddots&\ddots&\quad\vdots\\
\vdots &\ddots&\Lambda&\quad0
\\
\frac{1}{(s_\nu-1)!}\frac{\text{d}^{s_\nu-1}}{\text{d}k^{s_\nu-1}}
\Lambda&\ldots&\frac{1}{1!}\frac{\text{d}}{\text{d}k}\Lambda&\quad\Lambda
\end{array}\right),\quad
{\bf\Omega}_\nu \equiv \left(\begin{array}{cccc}
\Omega&0 & \ldots&\quad 0\\
\frac{1}{1!}\frac{\text{d}}{\text{d}k}\Omega
&\ddots&\ddots&\quad\vdots\\
\vdots &\ddots&\Omega&\quad0
\\
\frac{1}{(s_\nu-1)!}\frac{\text{d}^{s_\nu-1}}{\text{d}k^{s_\nu-1}}
\Omega&\ldots&\frac{1}{1!}\frac{\text{d}}{\text{d}k}\Omega&\quad\Omega
\end{array}\right),
\label{S29}\end{equation}
\begin{equation}
{\bf U}_\nu \equiv \left(\begin{array}{cccc}
 U&0 & \ldots& \quad 0\\
\frac{1}{1!}\frac{\text{d}}{\text{d}k} U
&\ddots&\ddots&\quad\vdots\\
\vdots &\ddots& U&\quad 0
\\
\frac{1}{(s_\nu-1)!}\frac{\text{d}^{s_\nu-1}}{\text{d}k^{s_\nu-1}}
 U&\ldots&\frac{1}{1!}\frac{\text{d}}{\text{d}k} U&\quad U
\end{array}\right),\quad
{\bf V}_\nu \equiv \left(\begin{array}{cccc}
 V&0 & \ldots&\quad 0\\
\frac{1}{1!}\frac{\text{d}}{\text{d}k} V
&\ddots&\ddots&\quad\vdots\\
\vdots &\ddots& V&\quad0
\\
\frac{1}{(s_\nu-1)!}\frac{\text{d}^{s_\nu-1}}{\text{d}k^{s_\nu-1}}
 V&\ldots&\frac{1}{1!}\frac{\text{d}}{\text{d}k} V&\quad V
\end{array}\right).
\end{equation}
Indeed, this is due to the fact that the matrix multiplication in (\ref{S28})
exactly reproduces the Leibniz rule for higher-order derivatives of a product.
Similarly, using the equations for $\Gamma^{-1}$,
\begin{subequations}
\begin{eqnarray}
\partial_x \Gamma^{-1}(k,x,t) &=& -\Lambda(k)\Gamma^{-1}(k,x,t) -
\Gamma^{-1}(k,x,t)U(k,x,t), \\
\partial_t \Gamma^{-1}(k,x,t) &=& -\Omega(k)\Gamma^{-1}(k,x,t) -
\Gamma^{-1}(k,x,t)V(k,x,t),\end{eqnarray}
\end{subequations}
one finds that
\begin{subequations}
\label{S30}
\begin{eqnarray}
\partial_x \overline{\bf\Gamma}_\nu(k,x,t) &=&
-\overline{\bf\Lambda}_\nu(k)\overline{\bf\Gamma}_\nu(k,x,t)
 - \overline{\bf\Gamma}_\nu(k,x,t)\overline{\bf U}_\nu(k,x,t),
\label{S30a}\\
\partial_t \overline{\bf\Gamma}_\nu(k,x,t) &=&
-\overline{\bf\Omega}_\nu(k)\overline{\bf\Gamma}_\nu(k,x,t)
 - \overline{\bf\Gamma}_\nu(k,x,t)\overline{\bf V}_\nu(k,x,t),
\label{S30b}\end{eqnarray}
\end{subequations}
where $\overline{\bf\Lambda}_\nu$,
$\overline{\bf\Omega}_\nu$, $\overline{\bf U}_\nu$, and $\overline{\bf V}_\nu$
are upper-triangular block-Toeplitz matrices:
\begin{equation}
\overline{\bf\Lambda}_\nu = \left(\begin{array}{cccc}
\Lambda\;&\frac{1}{1!}\frac{\text{d}}{\text{d}k}\Lambda&\ldots&
\frac{1}{(s_\nu-1)!}\frac{\text{d}^{s_\nu-1}}{\text{d}k^{s_\nu-1}}\Lambda\\
0\;
&\Lambda&\ddots& \vdots \\
\vdots\; &\ddots&\ddots&\frac{1}{1!}\frac{\text{d}}{\text{d}k}\Lambda
\\
0\; &\ldots&0 & \Lambda
\end{array}\right),\quad
\overline{\bf\Omega}_\nu = \left(\begin{array}{cccc} \Omega\;
&\frac{1}{1!}\frac{\text{d}}{\text{d}k}\Omega&\ldots&
\frac{1}{(s_\nu-1)!}\frac{\text{d}^{s_\nu-1}}{\text{d}k^{s_\nu-1}}\Omega\\
0\;
&\Omega&\ddots& \vdots \\
\vdots\; &\ddots&\ddots&\frac{1}{1!}\frac{\text{d}}{\text{d}k}\Omega \\
0\; &\ldots & 0  &\Omega
\end{array}\right),
\label{S31}\end{equation}
\begin{equation}
\overline{\bf U}_\nu = \left(\begin{array}{cccc}
 U\;&\frac{1}{1!}\frac{\text{d}}{\text{d}k} U&\ldots&
\frac{1}{(s_\nu-1)!}\frac{\text{d}^{s_\nu-1}}{\text{d}k^{s_\nu-1}} U\\
0\;
& U&\ddots& \vdots \\
\vdots\; &\ddots&\ddots&\frac{1}{1!}\frac{\text{d}}{\text{d}k} U\\
0\; &\ldots&0 &  U
\end{array}\right),\quad
\overline{\bf V}_\nu = \left(\begin{array}{cccc}
 V\; &\frac{1}{1!}\frac{\text{d}}{\text{d}k} V&\ldots&
\frac{1}{(s_\nu-1)!}\frac{\text{d}^{s_\nu-1}}{\text{d}k^{s_\nu-1}} V\\
0\;
& V&\ddots&\vdots \\
\vdots\; &\ddots&\ddots&\frac{1}{1!}\frac{\text{d}}{\text{d}k} V\\
0\; &\ldots &0 &  V
\end{array}\right).
\end{equation}
To obtain the $(x,t)$-dependence of the $p$-vectors, we differentiate
equations (\ref{S14}) and (\ref{S15}). Utilizing Eqs. (\ref{S29}) and
(\ref{S30}), we find that
\begin{equation} \label{Gammapx}
{\bf \Gamma}_\nu(\kappa_\nu)\left\{\left[\partial_x+{\bf \Lambda}_\nu(\kappa_\nu)\right]
\left(\begin{array}{c} |p^{(\nu)}_1\rangle \\ \vdots \\
|p^{(\nu)}_{s_\nu}\rangle
\end{array}\right) \right\}=0,
\end{equation}
and
\begin{equation} \label{Gammapt}
{\bf \Gamma}_\nu(\kappa_\nu)\left\{\left[\partial_t+
{\bf \Omega}_\nu(\kappa_\nu)\right]
\left(\begin{array}{c} |p^{(\nu)}_1\rangle \\ \vdots \\
|p^{(\nu)}_{s_\nu}\rangle
\end{array}\right) \right\}=0.
\end{equation}
Due to the invariance properties (see explanations below), we can set the quantities
inside the curly brackets of Eqs. (\ref{Gammapx}) and (\ref{Gammapt}) to be zero
without any loss of generality:
\begin{equation} \label{peq}
\left[\partial_x+{\bf \Lambda}_\nu(\kappa_\nu)\right]
\left(\begin{array}{c} |p^{(\nu)}_1\rangle \\ \vdots \\
|p^{(\nu)}_{s_\nu}\rangle
\end{array}\right)=0,
\quad \quad
\left[\partial_t+
{\bf \Omega}_\nu(\kappa_\nu)\right]
\left(\begin{array}{c} |p^{(\nu)}_1\rangle \\ \vdots \\
|p^{(\nu)}_{s_\nu}\rangle
\end{array}\right)=0.
\end{equation}
The reason for it is the uniqueness  of solution to the Riemann-Hilbert problem for
the given spectral data. Thus, the $(x, t)$-evolution of $|p\rangle$ vectors is
\begin{equation}
\left(\begin{array}{c} |p^{(\nu)}_1\rangle \\ \vdots \\
|p^{(\nu)}_{s_\nu}\rangle\end{array}\right) = \exp\left\{{-\bf\Lambda}_\nu(\kappa_\nu)x
-{\bf\Omega}_\nu(\kappa_\nu)t\right\}
\left(\begin{array}{c} |p^{(\nu)}_{01}\rangle \\ \vdots \\
|p^{(\nu)}_{0s_\nu}\rangle\end{array}\right).
\label{S32a}
\end{equation}
By similar arguments, the $(x, t)$-evolution of $\langle \overline{p}|$ vectors is
\begin{equation}
(\langle \overline{p}^{(\nu)}_1|,\ldots,\langle \overline{p}^{(\nu)}_{s_\nu}|) =
(\langle \overline{p}^{(\nu)}_{01}|,\ldots,\langle \overline{p}^{(\nu)}_{0s_\nu}|)
\exp\left\{\overline{\bf\Lambda}_\nu(\overline{\kappa}_\nu)x
+\overline{\bf\Omega}_\nu(\overline{\kappa}_\nu)t\right\}.
\label{S32b}
\end{equation}
Here the subscript ``0''  is used to denote constant vectors.
The exponential functions in the above two equations can be readily determined.
Indeed, by using the property that the operation of
raising to the exponent of a diagonal matrix (such as $\Lambda(k)x+\Omega(k)t$ here)
commutes with the construction of the block-Toeplitz matrix (see appendix in Ref.~\cite{SIAM}),
we find that
\begin{subequations}
\label{S33}
\begin{equation}
\exp\left\{{-\bf\Lambda}_\nu(\kappa_\nu)x -{\bf\Omega}_\nu(\kappa_\nu)t\right\} =
\left(\begin{array}{cccc}
E(k_1)&0 & \ldots&\; 0\\
\frac{1}{1!}\frac{\text{d}}{\text{d}k}E(k_1)
& \ddots&\ddots&\;\vdots\\
\vdots &\ddots&E(k_1)&\;
0 \\
\frac{1}{(s_\nu-1)!}\frac{\text{d}^{s_\nu-1}}{\text{d}k^{s_\nu-1}}E(k_1)&\ldots&
\frac{1}{1!}\frac{\text{d}}{\text{d}k}E(k_1)&\; E(k_1)\end{array}\right),
\label{S33a}\end{equation}
and
\begin{equation}
\exp\left\{\overline{\bf\Lambda}_\nu(\overline{\kappa}_\nu)x
+\overline{\bf\Omega}_\nu(\overline{\kappa}_\nu)t\right\}
 = \left(\begin{array}{cccc}
E^{-1}(\overline{k}_1)&\frac{1}{1!}\frac{\text{d}}{\text{d}k}E^{-1}(\overline{k}_1)&\ldots&
\frac{1}{(s_\nu-1)!}\frac{\text{d}^{s_\nu-1}}{\text{d}k^{s_\nu-1}}E^{-1}(\overline{k}_1)\\
0&E^{-1}(\overline{k}_1)&\ddots&
\vdots \\
\vdots&\ddots&\ddots&\frac{1}{1!}\frac{\text{d}}{\text{d}k}E^{-1}(\overline{k}_1)\\
0&\ldots&0 &E^{-1}(\overline{k}_1)
\end{array}\right),
\label{S33b} \end{equation}
\end{subequations}
where $E(k) \equiv \exp\left\{-\Lambda(k)x -\Omega(k)t\right\}$. After the spatial
and temporal evolutions of vectors $|p\rangle$ and $\langle \overline{p}|$ have
been given from Eqs. (\ref{S32a}) to (\ref{S33}), the soliton matrices
(\ref{mainrepr}) are then obtained. Eventually, the soliton solutions are derived
from Eq. (\ref{RH3}) by taking the limit $k \to \infty$. For the three-wave
interaction model, soliton solutions are given by Eqs. (\ref{UV3wave}) and
(\ref{u1u2u3}). The corresponding eigenfunctions of the $N$-dimensional
Zakharov-Shabat spectral problem with those soliton (reflection-less) potentials
are simply the column vectors of the soliton matrices $\Gamma(k)$ and
$\Gamma^{-1}(k)$ in (\ref{mainrepr}) by taking $k$ to be zeros ($k_j$,
$\overline{k}_j$) (which give discrete eigenfunctions) and with $k$ lying on the
real axis (which give continuous eigenfunctions).

Lastly, we explain why other $|p\rangle$ solutions to Eqs. (\ref{Gammapx}) and
(\ref{Gammapt}) give the same soliton matrices as those from Eq. (\ref{peq}).
Notice that equations (\ref{Gammapx}) for all $\nu$ blocks can be written
in the following compact form:
\begin{equation}\label{peq2}
{\bf \Gamma_{\bf B}} \left(\partial_x+
{\bf \Omega}_{\bf B}\right)
\left(\begin{array}{c} |p_1^{(1)}\rangle \\ \vdots \\ |p_{s_1}^{(1)}\rangle \\ \vdots \\
|p_1^{(r_1)}\rangle \\ \vdots  \\ |p_{s_{r_1}}^{(r_1)}\rangle \end{array}\right)
=0,
\quad \quad\quad
{\bf \Omega_{\bf B}} \equiv \left(\begin{array}{ccc}{\bf \Omega}_1(\kappa_1) & \null & 0\\
\null & \ddots & \null \\ 0 & \null & {\bf \Omega}_{r_1}(\kappa_{r_1}) \end{array}\right).
\end{equation}
According to the invariance properties in the subsection \ref{invarian},
any two vectors in the kernel of matrix ${\bf \Gamma_{\bf B}}$ are linearly
dependent. Thus the most general $|p\rangle$ solutions to Eq. (\ref{Gammapx}) are such that
\begin{equation}
\left(\partial_x+{\bf \Omega}_{\bf B}\right)
\left(\begin{array}{c} |\widetilde{p}_1^{(1)}\rangle \\ \vdots \\ |\widetilde{p}_{s_1}^{(1)}\rangle \\ \vdots \\
|\widetilde{p}_1^{(r_1)}\rangle \\ \vdots  \\ |\widetilde{p}_{s_{r_1}}^{(r_1)}\rangle \end{array}\right)
=B^T(x, t)
\left(\begin{array}{c} |\widetilde{p}_1^{(1)}\rangle \\ \vdots \\ |\widetilde{p}_{s_1}^{(1)}\rangle \\ \vdots \\
|\widetilde{p}_1^{(r_1)}\rangle \\ \vdots  \\ |\widetilde{p}_{s_{r_1}}^{(r_1)}\rangle \end{array}\right),
\end{equation}
where $B$ is an invariance matrix which depends on $x$ and $t$ in general [see Eq. (\ref{ptilde})].
To show that these $|\widetilde{p}\rangle$ vectors give the same soliton matrices (\ref{mainrepr})
as the $|p\rangle$ vectors from Eq. (\ref{peq}), we define a matrix function $G(x, t)$
which satisfies the following differential equation and initial condition:
\[
\partial_x G(x, t)=B^T(x, t)G(x, t), \quad \quad G|_{x=0}=I.
\]
Because the matrix $B$ here is an invariance matrix and $G(x=0)=I$,
obviously the function $G(x, t)$ is an invariance matrix as well
(note that $G$ is always non-degenerate from its construction).
In addition, $G^{-1}$ is also an invariance matrix.
Now for any solution $|\widetilde{p}\rangle$ of Eq. (\ref{peq2}), we define new
vectors $|p\rangle$ as
\[
\left(\begin{array}{c} |p_1^{(1)}\rangle \\ \vdots \\ |p_{s_1}^{(1)}\rangle \\ \vdots \\
|p_1^{(r_1)}\rangle \\ \vdots  \\ |p_{s_{r_1}}^{(r_1)}\rangle \end{array}\right)
=G^{-1}
\left(\begin{array}{c} |\widetilde{p}_1^{(1)}\rangle \\ \vdots \\ |\widetilde{p}_{s_1}^{(1)}\rangle \\ \vdots \\
|\widetilde{p}_1^{(r_1)}\rangle \\ \vdots  \\ |\widetilde{p}_{s_{r_1}}^{(r_1)}\rangle \end{array}\right).
\]
Then these $|p\rangle$ vectors satisfy the first equation in (\ref{peq}).
This can be checked directly
by substituting the above equation into (\ref{peq}) and noting that
matrices $G$ and ${\bf \Omega}_{\bf B}$ commute by virtue of Eq. (\ref{comm1})
and the fact that matrices ${\bf \Omega}_{\bf B}$ and ${\bf \Gamma}_{\bf B}$
have identical forms.
Since $G^{-1}$ is an invariance matrix, $|p\rangle$ and $|\widetilde{p}\rangle$ vectors
as related above naturally give the same soliton matrices (\ref{mainrepr}).
Thus there is no any loss of generality by picking solutions $|p\rangle$ of Eq. (\ref{Gammapx})
such that the first equation in (\ref{peq}) holds.
By the same argument, there is no loss of generality by
picking solutions $|p\rangle$ of Eq. (\ref{Gammapt})
such that the second equation in (\ref{peq}) holds.

\section{Applications to the three-wave interaction system}
\label{sec3wave}

To illustrate the above general results, we apply them to the three-wave interaction
model (\ref{3wave}) and display various higher-order soliton solutions. In this case,
the involution property (\ref{invol}) holds, thus all zeros are normal and
appear in complex conjugate pairs.
The soliton matrix $\Gamma(k)$ is given by Eq. (\ref{mainrepr}a), where
$\langle \overline{p}|=|p\rangle^\dagger$, and the $(x, t)$-evolution
of $|p\rangle$ vectors is given by Eqs. (\ref{S32a}) and (\ref{S33}a).
The general higher-order soliton solutions of the three-wave system
are then given by Eq. (\ref{u1u2u3}), where
\begin{equation}\label{Gamma1}
\Phi^{(1)}=\Gamma^{(1)}=-
(|p^{(1)}_1\rangle,\ldots,|p^{(1)}_{s_1}\rangle,\ldots,|p^{(r_1)}_1\rangle,\ldots,|p^{(r_1)}_{s_{r_1}}\rangle)
\overline{\cal K}{}^{-1}\left(\begin{array}{c}
\langle\overline{p}^{(1)}_1|\\ \vdots\\ \langle\overline{p}^{(1)}_{s_1}| \\
\vdots \\ \langle\overline{p}^{(r_1)}_1| \\ \vdots \\
\langle\overline{p}^{(r_1)}_{s_{r_1}}|\end{array}\right),
\end{equation}
and matrix $\overline{\cal K}$ is given in Eq. (\ref{KKbar}).
In all our solutions, we fix the parameters in the dispersion laws (\ref{disper}) as
$(a_1, a_2, a_3)=(1, 0.5, -0.5)$ and $(b_1, b_2, b_3)=(1, 1.5, 0.5)$.

\subsection{Soliton solutions for a single pair of non-elementary zeros}
First, we derive soliton solutions corresponding to a single pair of non-elementary
zeros. In particular, we consider the rank sequence \{1, 2\} of a pair of zeros
$(k_1, \overline{k}_1)$. In this case, $r_1=2$ and $r_2=1$. Using formula
(\ref{Nfree}) (for the case of involution)   we get the number of free complex
parameters in the soliton solution:
\[
{\cal N}_\mathrm{free} = 3(2+1) + 1 - (4 + 1) = 10 -5 =5.
\]
There are three $|p\rangle$ vectors, $|p_1^{(1)}\rangle, |p_2^{(1)}\rangle$ and
$|p_1^{(2)}\rangle$ in Eq. (\ref{Gamma1}). When $k_1$ and the initial values
$[|p_{01}^{(1)}\rangle, |p_{02}^{(1)}\rangle, |p_{01}^{(2)}\rangle]$ of these
vectors are provided, the soliton solutions
(\ref{u1u2u3}) will then be completely determined.

In the present case, the block sequence reads $\{s_1, s_2\}=\{2, 1\}$, and the invariance
matrix $B$ for this case can be readily obtained from the general formula
(\ref{Bform}) as
\[B=\left(\begin{array}{ccc} b_{11} & b_{12}& b_{13} \\
                     0 & b_{11} & 0 \\
                     0 & b_{32} & b_{33} \end{array}\right),\]
which indeed has five free complex parameters [see Eq. (\ref{number})]. The
invariance matrix $\overline{B}$ is just the Hermitian of the $B$ matrix.

To display these soliton solutions, we choose $k_1=1+i$, $|p_{02}^{(1)}\rangle=[-1,
i, 1-i]^T$, $|p_{01}^{(2)}\rangle=[1, 0.5, -1]^T$. When $|p_{01}^{(1)}\rangle=[1,
1+i, 0.5]^T$ (the generic case), the solutions are plotted in the top row of Fig.
1. In two non-generic cases (where some elements of the $|p\rangle$ vectors
vanish), $|p_{01}^{(1)}\rangle=[0, 1+i, 0.5]^T$ and $|p_{01}^{(1)}\rangle=[1, 0,
0.5]^T$, the solutions are plotted in the second and third rows of Fig. 1
respectively. We see that in the generic case, three sech waves in the three
components interact and then separate into the same sech waves with their positions
shifted. In other words, this is a $u_1 (\mbox{sech}) + u_2 (\mbox{sech}) +u_3
(\mbox{sech}) \rightarrow u_1 (\mbox{sech}) + u_2 (\mbox{sech}) +u_3 (\mbox{sech})$
process. What happens is that the initial pumping ($u_3$) wave breaks up into two
sech waves in the other two components ($u_1$ and $u_2$), while simultaneously the
two initial $u_1$ and $u_2$ waves combine into a pumping sech wave. Thus this
process is a combination of two sub-processes: $u_3 \rightarrow u_1+u_2$ and
$u_1+u_2 \rightarrow u_3$. This phenomenon seems related to the rank sequence \{1,
2\} of the present solitons and the fact that, the rank sequence \{1\} itself
describes the breakup of a pumping sech wave into two non-pumping sech waves, while
the rank sequence \{2\} itself describes the reserve process. In the non-generic
cases, these solutions can describe the $u_1 (\mbox{sech}) + u_2
(\mbox{second-order}) \rightarrow u_2 (\mbox{sech}) + u_3 (\mbox{sech})$ process,
the $u_1 (\mbox{sech}) + u_2 (\mbox{sech}) +u_3 (\mbox{sech}) \rightarrow u_3
(\mbox{second-order})$ process (see Fig. 1, second and third rows), and many
others. In the solutions of Fig. 1, the $a_j$ and $b_j$ parameters are such that
$u_2 < u_3 < u_1$. If $u_1 < u_3 < u_2$, the processes will be exactly the opposite
(see \cite{SIAM}). Thus our solutions can describe the opposite processes of Fig. 1
as well.

\subsection{Soliton solutions for two pairs of simple zeros}
Here we derive soliton solutions corresponding to two pairs of simple zeros in the
three-wave system (\ref{3wave}). Some solutions belonging to this category have
been presented in \cite{3wavezakharov,3wave3}. But we will show that those
solutions are only special (non-generic) solutions for two pairs of simple zeros.
Below, the more general solutions for this case will be presented.

In this case, $r_1^{(1)}=r_1^{(2)}=1$. By using formula (\ref{Nfree}) for the case of
involution (\ref{invol}) and with two pairs of zeros,
we readily obtain that the number of free complex parameters in the solution is 6:
\[
{\cal N}_\mathrm{free}= 2(3\times1 + 1 - 1) = 6.
\]
Indeed, there are two $|p\rangle$ vectors in Eq. (\ref{Gamma1}). Together with the
two zeros $k_1$ and $k_2$, there are 8 complex parameters in the soliton solutions.
However, the $2\times 2$ invariance matrix $B$ in this case is diagonal and has two
free (diagonal) complex parameters.

Three solutions, with $k_1=1+i, k_2=-1+0.5i$ and
three different sets of $|p_{01}^{(1)}\rangle$ and $|p_{01}^{(2)}\rangle$ vectors,
are displayed in Fig. 2.
In the generic case where
$|p_{01}^{(1)}\rangle=[1, 1+i, 0.5]^T$ and $|p_{01}^{(2)}\rangle=[1, 0.5, -1]^T$ (see top row of Fig. 2),
the solution describes the breakup of a higher-order pumping ($u_3$) wave
into two higher-order $u_1$ and $u_2$ waves. This is analogous to
solutions for a single pair of elementary zeros with algebraic multiplicity 2
(see \cite{SIAM}). In the non-generic case where
$|p_{01}^{(1)}\rangle=[0, 1+i, 0.5]^T$ and $|p_{01}^{(2)}\rangle=[1, 0.5, -1]^T$
(second row in Fig. 2), the present solutions
can describe the $u_2 (\mbox{sech}) + u_3 (\mbox{sech}) \rightarrow u_1 (\mbox{sech}) + u_2 (\mbox{second-order})$ process.
This process has been seen in \cite{SIAM} for elementary zeros as well.
More interestingly, in the non-generic case when
$p_{01}^{(1)}[1]=p_{01}^{(2)}[3]=0$, these solutions describe the elastic interaction of
a sech $u_1$ wave with a sech $u_2$ wave (see bottom row of Fig. 2).
These are precisely the soliton solutions presented in \cite{3wavezakharov,3wave3}.
We see that these solutions are simply non-generic solutions for two pairs of simple zeros.

\subsection{Soliton solutions for two pairs of higher-order zeros}
Lastly, we consider two pairs of zeros, one simple and the other one elementary
with the algebraic multiplicity 2. Let us say $k_1$ is the elementary zero, and
$k_2 \:(\ne k_1) $ is the simple zero.
Then the rank sequence for $k_1$ is \{1, 1\}, and
the rank sequence for $k_2$ is $\{1\}$.
Thus, $r_1^{(1)}=1$, $r_2^{(1)}=1$, and
$r_1^{(2)}=1$. By formula (\ref{Nfree}) we have
\[
{\cal N}_\mathrm{free} = 3(1+1) +1 - (1+1) + 3\times1 +1 - 1 = 8.
\]
Indeed, in this case $s^{(1)}_1=2$ and $s^{(2)}_1=1$, hence there are 11 complex
parameters in the soliton solutions (9 in the three $|p\rangle$ vectors, plus the
two zeros $k_1$ and $k_2$). The invariance matrix $B$ can be found from the general
formula (\ref{Bform}) as
\[B=\left(\begin{array}{ccc} b_{11} & b_{12}& 0 \\
                     0 & b_{11} & 0 \\
                     0 & 0 & b_{33} \end{array}\right),\]
which has three free complex parameters. Thus ${\cal N}_\mathrm{free}=11-3=8$
as calculated above.

Three solutions, with $k_1=1+i, k_2=-1+0.5i$, $|p_{02}^{(1)}\rangle=[-1, i,
1-i]^T$,  and three different sets of $|p_{01}^{(1)}\rangle$ and
$|p_{01}^{(2)}\rangle$ vectors, are displayed in Fig. 3. In the generic case (first
row in Fig. 3), this solution describes the breakup of a higher-order pumping wave
($u_3$) into the other $u_1$ and $u_2$ components (both higher-order). In
non-generic cases, it can describe processes such as $u_2 (\mbox{sech}) +u_3
(\mbox{higher-order}) \rightarrow u_1 (\mbox{higher-order}) + u_2
(\mbox{higher-order})$  (second row of Fig. 3), $u_1 (\mbox{sech}) + u_2
(\mbox{sech}) + u_3 (\mbox{sech}) \rightarrow u_1 (\mbox{higher-order}) + u_2
(\mbox{higher-order})$ (last row of Fig. 3), and many others. The inverse processes
of Fig. 3 can also be described by choosing $a_j$ and $b_j$ values such that $u_1 <
u_3 < u_2$ instead of $u_2 < u_3 < u_1$ in Fig. 3.

\section{Conclusion and Discussion}

We have proposed a unified and systematic approach to study the higher-order
soliton solutions of nonlinear PDEs integrable by the $N\times N$-dimensional
Riemann-Hilbert problem. We have derived the complete solution to
the Riemann-Hilbert problem with an arbitrary number of higher-order zeros, and
characterized the discrete spectral data.
As a result, the most general forms of higher-order multi-soliton
solutions have been obtained in nonlinear PDEs integrable
through the $N\times N$-dimensional Riemann-Hilbert problem.
In other words, the most general reflection-less (soliton) potentials
in the $N$-dimensional Zakharov-Shabat operators have been derived.
The eigenfunctions associated with these reflection-less potentials
are readily available from our soliton matrices.
We have applied these general results to the three-wave interaction system,
and new higher-order soliton and two-soliton solutions have been presented.
These solutions reveal new processes such as $u_1+u_2+u_3 \leftrightarrow u_1+u_2+u_3$.
They also reproduce previous solitons in \cite{NMPZ84,3wavezakharov,3wave3,SIAM}
as special cases. Our results can be applied to
derive higher-order multi-solitons in the NLS equation and
the Manakov equations as well, but this is not pursued in this article.

The results obtained in this paper are significant from both physical and
mathematical points of view. Physically, our results completely characterized
higher-order solitons and multi-solitons in important physical systems such
as the three-wave interaction equation, the NLS equation and the Manakov equations.
These higher-order solitons can describe new physical processes such as
those displayed in Figs. 1 - 3. If these integrable equations are perturbed
(which is inevitable in a real-world problem), our higher-order solitons then
become the starting point for the development of a soliton-perturbation theory
which could determine what happens to these higher-order solitons under external
or internal perturbations \cite{hasegawa,Yang00}.
From the mathematical point of view, our results completely characterized
the discrete spectral data of higher-order zeros
in a general $N$-dimensional Riemann-Hilbert problem.
These results will be useful for many purposes such as proving the completeness
of eigenfunctions in a $N$-dimensional Zakharov-Shabat spectral problem
with arbitrary localized potentials. The difficulty of such a proof is caused
by higher-order zeros. With our results, this difficulty can be hopefully removed.

From a broader perspective, our results are closely related to many other physical and mathematical
problems. For instance, the lump solutions in the Kadomtsev-Petviashvili~I equation
are given by the higher-order poles of the time-dependent Schr\"odinger equation.
In \cite{ablowitz, ablowitz2}, lump solutions corresponding to certain special
higher-order poles were derived, but the most general lump solutions still remain an open
question. Note that the time-dependent Schr\"odinger equation is an infinite-dimensional
system compared to our present $N$-dimensional Riemann-Hilbert system. But
the ideas used in this paper might be generalizable to the
time-dependent Schr\"odinger equation as well. This remains to be seen.

\section{Acknowledgments}

This work was supported in part by NASA and AFOSR grants.
J.Y. acknowledges helpful discussions with Jonathan Sands.
V.S. would like to thank the University of Vermont for support of his visit during
which this work was done.

\appendix
\section{General Riemann-Hilbert problem with abnormal zeros}
\label{appendix1}

Here we show that our soliton matrices of section~\ref{multpl} can be
generalized to the case of Riemann-Hilbert problem with abnormal zeros. However,
due to the lack of important applications, we will show a simple example, which
corresponds to a pair of zeros with different geometric multiplicities but
the same algebraic multiplicity. Then we
comment on the general case of several non-paired zeros.

Let us use the simplest example to show the idea behind generalization of our
results to the general Riemann-Hilbert problem with abnormal zeros. Consider one pair
of zeros $(k_1,\overline{k}_1)$ which have the same algebraic multiplicity 2 but
different geometric multiplicities which are 1 and 2 respectively.
The corresponding soliton matrices are as follows:
\begin{equation}
\Gamma(k) = I  + \frac{(\overline{k}_1-k_1)\left(|v_1\rangle\langle\overline{v}_1|
+ |v_2\rangle\langle\overline{v}_2|\right)}{k - \overline{k}_1},
\label{A1a}\end{equation}
\begin{equation}
\Gamma^{-1}(k) = \left(I + \frac{(k_1 -
\overline{k}_1)|v_1\rangle\langle\overline{v}_1|}{k - k_1}\right) \left(I +
\frac{(k_1 - \overline{k}_1)|v_2\rangle\langle\overline{v}_2|}{k - k_1}\right),
\label{A1b}\end{equation}
with the conditions that $\langle\overline{v}_j|v_j\rangle=1$,
$\langle\overline{v}_2|v_1\rangle=0$ and $\langle\overline{v}_1|v_2\rangle\ne0$. To
verify that the above matrices are indeed inverse to each other it is enough to
rewrite the matrix $\Gamma(k)$ in the form
\begin{equation}
\Gamma(k) = \left(I +
\frac{(\overline{k}_1-k_1)|v_2\rangle\langle\overline{v}_2|}{k -
\overline{k}_1}\right) \left(I +
\frac{(\overline{k}_1-k_1)|v_1\rangle\langle\overline{v}_1|}{k-\overline{k}_1}\right)
\label{A2}\end{equation}
and take into account that $P_j\equiv|v_j\rangle\langle\overline{v}_j|$ is a
projector. Equations (\ref{A1b}) and (\ref{A2}) are in fact the product
representations of the form (\ref{S6}). Now let us show that there are exactly two
solutions to $\langle\overline{p}|\Gamma^{-1}(\overline{k}_1)=0$. Indeed, the
corresponding null vectors are as follows
\begin{equation}
\langle\overline{p}_1| = \langle\overline{v}_1|,\quad \langle\overline{p}_2| =
\langle\overline{v}_2|.
\label{A3}\end{equation}
This is due to the fact that $\Gamma^{-1}(\overline{k}_1)=(I-P_1)(I-P_2)$. But on
the other hand  there is just one solution to $\Gamma(k_1)|p\rangle=0$:
$|p_1\rangle = |v_1\rangle$. Suppose that there is another solution $|p_2\rangle$
to $\Gamma(k_1)|p\rangle=0$ linearly independent from $|p_1\rangle$. We have then
using formula (\ref{A1a}) for $\Gamma(k_1)$:
\begin{equation}
|p_2\rangle = |v_1\rangle\langle\overline{v}_1|p_2\rangle +
|v_2\rangle\langle\overline{v}_2|p_2\rangle.
\label{A5}\end{equation}
Thus $|p_2\rangle = a|v_1\rangle + b|v_2\rangle$. Using this in formula (\ref{A5})
we get, due to $\langle\overline{v}_2|v_1\rangle=0$ and
$\langle\overline{v}_1|v_2\rangle\ne0$,
\[
a|v_1\rangle\langle\overline{v}_1|v_2\rangle = 0,
\]
which is a contradiction, since $a\ne0$.

The soliton matrices given by formulae (\ref{A1a})-(\ref{A1b}) have the following
form in the standard notations of Lemma~\ref{lem1} of section~\ref{multpl}:
\begin{equation}
\Gamma(k) = I  + \frac{|\overline{q}_1\rangle\langle\overline{p}_2| +
|\overline{q}_2\rangle\langle\overline{p}_1|}{k - \overline{k}_1},
\label{A6}\end{equation}
\begin{equation}
\Gamma^{-1}(k) = I + \frac{|p_1\rangle\langle q_2| + |{p}_2\rangle\langle q_1|}{k -
k_1} + \frac{|p_1\rangle\langle q_1|}{(k - k_1)^2},
\label{A7}\end{equation}
where
\[
|\overline{q}_1\rangle = (\overline{k}_1-k_1)|v_2\rangle,\quad
|\overline{q}_2\rangle = (\overline{k}_1-k_1)|v_1\rangle,\quad \langle q_1| =
(k_1-\overline{k}_1)^2\langle\overline{v}_1|v_2\rangle\langle\overline{v}_2|,\quad
\langle q_2| = (k_1-\overline{k}_1)\langle\overline{v}_1|,
\]
\[
|{p}_2\rangle =
\frac{|v_2\rangle}{(k_1-\overline{k}_1)\langle\overline{v}_1|v_2\rangle}.
\]
Notice that $\Gamma(k)$ has two blocks of size 1, while $\Gamma^{-1}(k)$ has one
block of size 2. In general, for one pair of zeros with different geometric
multiplicities, the soliton matrices have the structure of Lemma~\ref{lem1} but
with different numbers of blocks in $\Gamma(k)$ and $\Gamma^{-1}(k)$, while the
total number of the $|p\rangle$- and $\langle\overline{p}|$-vectors appearing in
these matrices is the same and equals to the order of the pair of zeros.  One can
proceed to derive the representations similar to those in Lemma~\ref{lem4} for this
case. Evidently, due to the way of the derivation, the formulae will be similar
with the only difference in the number of blocks and block sizes in $\Gamma(k)$ and
$\Gamma^{-1}(k)$.

In the more general case of the Riemann-Hilbert problem with abnormal zeros, the zeros can
be non-paired (for instance, zero of order 2 in $C_+$ and two simple zeros in
$C_-$). Formally, this case can be obtained by ``splitting'' some of the zeros
inside pairs in several distinct zeros in the soliton matrices $\Gamma(k)$ and
$\Gamma^{-1}(k)$ discussed above, since this limit is obviously regular (the
geometric multiplicity of the zero to be split should be at least equal to the
number of the generated in this way new zeros, thus providing for the needed number
of blocks; formula (\ref{A6}), for instance, allows splitting of the zero
$k=\overline{k}_1$ of $\Gamma^{-1}(k)$ into two simple zeros). Thus, the most
general case can be handled starting from the case of just one pair of zeros, i.e.,
the case discussed above. The explicit expressions for the soliton matrices
$\Gamma(k)$ and $\Gamma^{-1}(k)$ will involve similar relations between the numbers
of zeros, their geometric multiplicities and the numbers and sizes of the
$\nu$-blocks of vectors as those in lemma~\ref{lem1}, though, obviously, with
different particular numbers for each of the two matrices.

\begin{figure}
\includegraphics{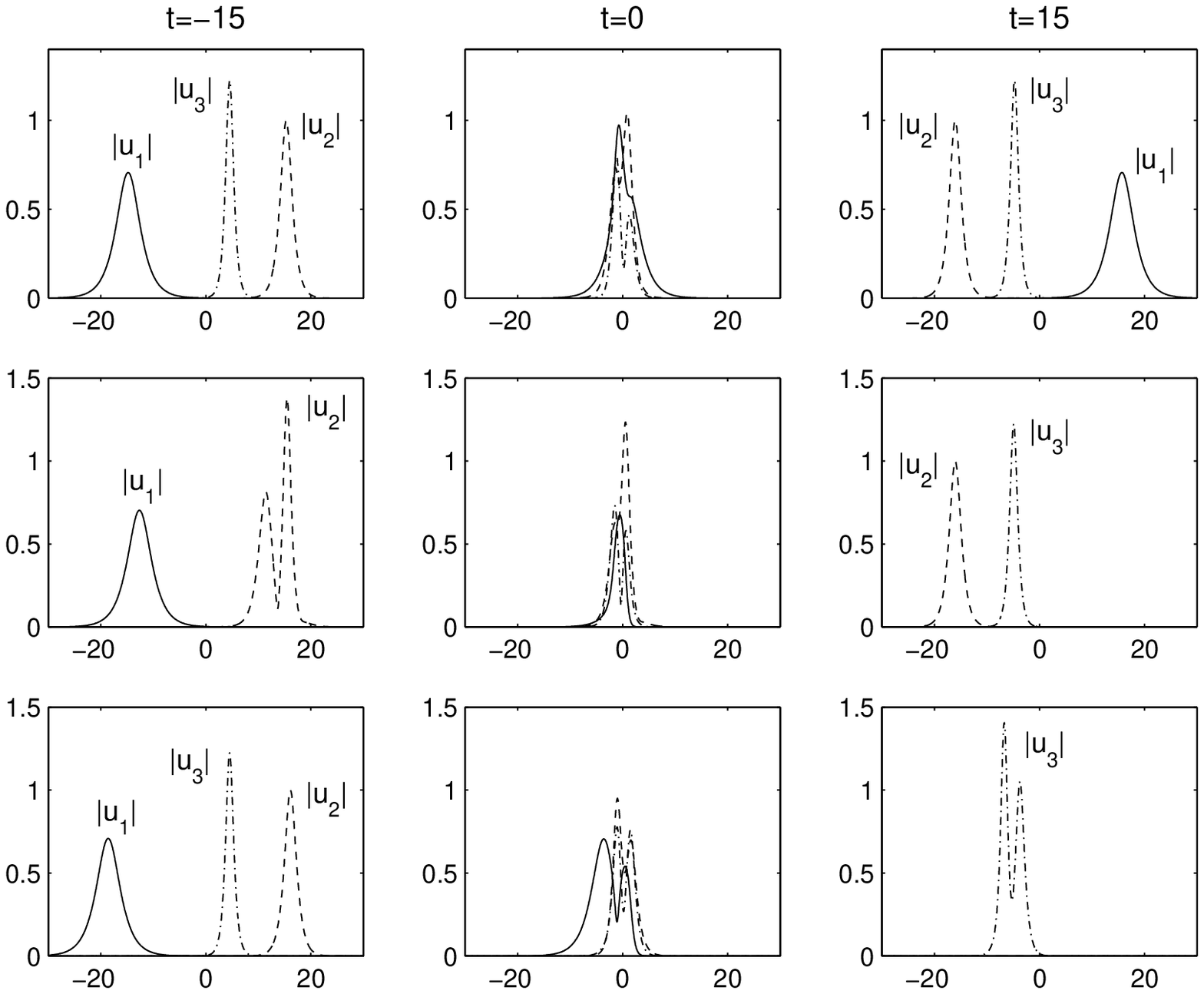}
\caption{Soliton solutions in the three-wave system (\ref{3wave}) corresponding to a single pair of zeros with rank sequence \{1, 2\} at time $t=-15, 0$ and 15.
Here, $k_1=1+i$, $|p_{02}^{(1)}\rangle=[-1, i, 1-i]^T$, $|p_{01}^{(2)}\rangle=[1, 0.5, -1]^T$.
First row: $|p_{01}^{(1)}\rangle=[1, 1+i, 0.5]^T$;
second row: $|p_{01}^{(1)}\rangle=[0, 1+i, 0.5]^T$;
third row: $|p_{01}^{(1)}\rangle=[1, 0, 0.5]^T$.  }
\end{figure}

\begin{figure}
\includegraphics{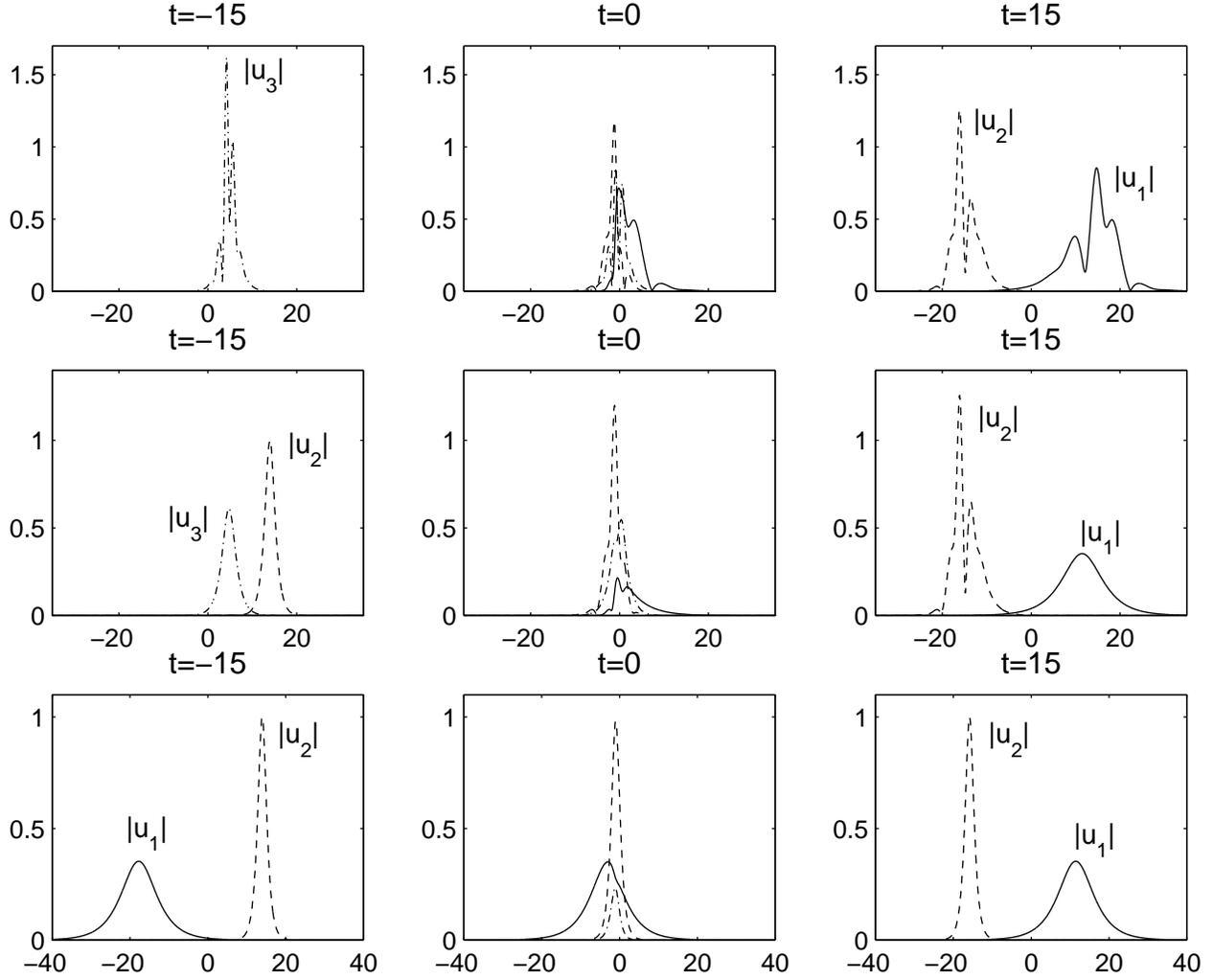}
\caption{Soliton solutions in the three-wave system (\ref{3wave}) corresponding to two pairs of simple zeros
at time $t=-15, 0$ and 15.
Here, $k_1=1+i$, $k_2=-1+0.5i$.
First row: $|p_{01}^{(1)}\rangle=[1, 1+i, 0.5]^T$, $|p_{01}^{(2)}\rangle=[1, 0.5, -1]^T$;
second row: $|p_{01}^{(1)}\rangle=[0, 1+i, 0.5]^T$, $|p_{01}^{(2)}\rangle=[1, 0.5, -1]^T$;
third row: $|p_{01}^{(1)}\rangle=[0, 1+i, 0.5]^T$, $|p_{01}^{(2)}\rangle=[1, 0.5, 0]^T$.  }
\end{figure}

\begin{figure}
\includegraphics{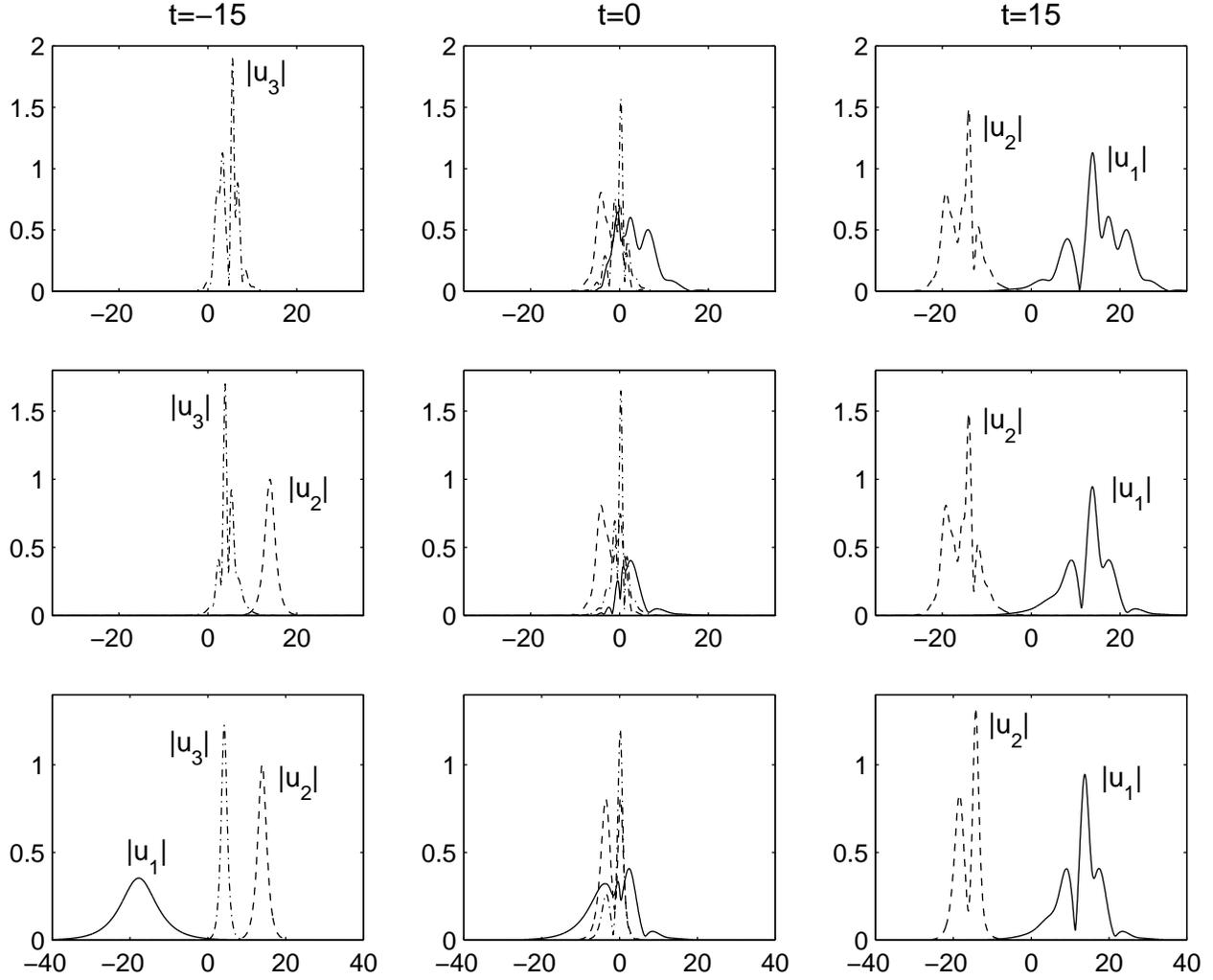}
\caption{Soliton solutions in the three-wave system (\ref{3wave}) corresponding to two
pairs of zeros --- one elementary with algebraic multiplicity 2, and the other one simple.
Here, $k_1=1+i$ (elementary zero), $k_2=-1+0.5i$ (simple zero),
and $|p_{02}^{(1)}\rangle=[-1, i, 1-i]^T$.
First row: $|p_{01}^{(1)}\rangle=[1, 1+i, 0.5]^T$, $|p_{01}^{(2)}\rangle=[1, 0.5, -1]^T$;
second row: $|p_{01}^{(1)}\rangle=[0, 1+i, 0.5]^T$, $|p_{01}^{(2)}\rangle=[1, 0.5, -1]^T$;
third row: $|p_{01}^{(1)}\rangle=[0, 1+i, 0.5]^T$, $|p_{01}^{(2)}\rangle=[1, 0.5, 0]^T$.  }

\end{figure}

\end{document}